\documentclass{JHEP3}
\usepackage{graphicx}
\usepackage{latexsym}

\def\PRL #1 #2 #3{{\em Phys. Rev. Lett. \/} {\bf#1} (#2) #3}
\def\NPB #1 #2 #3{{\em Nucl. Phys. \/} {\bf B#1} (#2) #3}
\def\NPBFS #1 #2 #3 #4{{\em Nucl. Phys. \/} {\bf B#2} [FS#1] (#3) #4}
\def\CMP #1 #2 #3{{\em Commun. Math. Phys. \/} {\bf #1} (#2) #3}
\def\PRD #1 #2 #3{{\em Phys. Rev. \/} {\bf D#1} (#2) #3}
\def\PLA #1 #2 #3{{\em Phys. Lett. \/} {\bf #1A} (#2) #3}
\def\PLB #1 #2 #3{{\em Phys. Lett. \/} {\bf B#1} (#2) #3}
\def\JMP #1 #2 #3{{\em J. Math. Phys. \/} {\bf #1} (#2) #3}
\def\PTP #1 #2 #3{{\em Prog. Theor. Phys. \/} {\bf #1} (#2) #3}
\def\SPTP #1 #2 #3{{\em Suppl. Prog. Theor. Phys. \/} {\bf #1} (#2) #3}
\def\AoP #1 #2 #3{{\em Ann. of Phys. \/} {\bf #1} (#2) #3}
\def\PNAS #1 #2 #3{{\em Proc. Natl. Acad. Sci. USA} {\bf #1} (#2) #3}
\def\RMP #1 #2 #3{{\em Rev. Mod. Phys. \/} {\bf #1} (#2) #3}
\def\PR #1 #2 #3{{\em Phys. Reports \/} {\bf #1} (#2) #3}
\def\AoM #1 #2 #3{{\em Ann. of Math. \/} {\bf #1} (#2) #3}
\def\UMN #1 #2 #3{{\em Usp. Mat. Nauk \/} {\bf #1} (#2) #3}
\def\FAP #1 #2 #3{{\em Funkt. Anal. Prilozheniya \/} {\bf #1} (#2) #3}
\def\FAaIA #1 #2 #3{{\em Functional Analysis and Its Application \/} {\bf
#1} (#2) #3}
\def\BAMS #1 #2 #3{{\em Bull. Am. Math. Soc. \/} {\bf #1} (#2)
#3} \def\TAMS #1 #2 #3{{\em Trans. Am. Math. Soc. \/} {\bf #1}
(#2) #3}
\def\InvM #1 #2 #3{{\em Invent. Math. \/} {\bf #1} (#2) #3}
\def\LMP #1 #2 #3{{\em Letters in Math. Phys. \/} {\bf #1} (#2) #3}
\def\IJMPA #1 #2 #3{{\em Int. J. Mod. Phys. \/} {\bf A#1} (#2) #3}
\def\AdM #1 #2 #3{{\em Advances in Math. \/} {\bf #1} (#2) #3}
\def\RMaP #1 #2 #3{{\em Reports on Math. Phys. \/} {\bf #1} (#2) #3}
\def\IJM #1 #2 #3{{\em Ill. J. Math. \/} {\bf #1} (#2) #3}
\def\APP #1 #2 #3{{\em Acta Phys. Polon. \/} {\bf #1} (#2) #3}
\def\TMP #1 #2 #3{{\em Theor. Mat. Phys. \/} {\bf #1} (#2) #3}
\def\JPA #1 #2 #3{{\em J. Physics \/} {\bf A#1} (#2) #3}
\def\JSM #1 #2 #3{{\em J. Soviet Math. \/} {\bf #1} (#2) #3}
\def\MPLA #1 #2 #3{{\em Mod. Phys. Lett. \/} {\bf A #1} (#2) #3}
\def\JETP #1 #2 #3{{\em Sov. Phys. JETP \/} {\bf #1} (#2) #3}
\def\JETPL #1 #2 #3{{\em  Sov. Phys. JETP Lett. \/} {\bf #1} (#2) #3}
\def\PHSA #1 #2 #3{{\em Physica} {\bf A#1} (#2) #3}
\def\CQG #1 #2 #3{{\em Class. Quantum Grav. \/} {\bf #1} (#2) #3}
\def\SJNP #1 #2 #3{{\em Sov. J. Nucl. Phys. (Yadern.Fiz.) \/} {\bf #1} (#2) #3}

\def\be{\begin{equation}}
\def\ee{\end{equation}}
\def\ba{\begin{array}} \def\ea{\end{array}}
\def\bea{\begin{eqnarray}}
\def\eea{\end{eqnarray}}

\title {Introduction to the Classical Theory of Higher Spins\footnote{A course of lectures given during
the year 2003--2004 at the Institute of Physics of Humboldt
University, Berlin (Germany), at the Advanced Summer School on
Modern Mathematical Physics, Dubna  (Russia), at the Department of
Theoretical Physics of Valencia University (Spain), Institute for
Theoretical Physics, NSC KIPT, Kharkov (Ukraine) and at the XIX-th
Max Born Symposium, Wroclaw University (Poland).}}

\author{
Dmitri Sorokin
\\
  Istituto Nazionale di Fisica Nucleare, Sezione di
Padova {\rm \&} Dipartimento di Fisica, Universit\`a degli Studi
di Padova, Via F. Marzolo 8, 35131 Padova, Italia
\\
and\\
Institute for Theoretical Physics, NSC KIPT, 61108 Kharkov, Ukraine
\\
 \email{}
 }

\author{
\\
\\
\email{}
 }

\abstract{We review main features and problems of higher spin
field theory and flash some ways along which it has been developed
over last decades.}


\preprint{DFPD 04/TH/10} 

\begin{document}

\section{Introduction}
Several dozens of years of intensive study, which involved enormous
amount of theoretical brain power, have resulted in a deep insight
into various fundamental features of String Theory, and every new
year of research brings us new and new aspects of its immense
structure. By now, for example, we know in detail low energy
field--theoretical limits of String Theory which correspond to
massless excitations over different string vacua. These are
described by ten--dimensional supergravities whose supermultiplets
consist of fields of spin not higher than two and higher order
corrections thereof. This region of String Theory also contains
various types of branes which reflect dualities between different
string vacua. The classical dynamics of the supergravity fields is
well known from the analysis of the classical supergravity actions
and equations of motion which possess an interesting geometrical and
symmetry structure based on supersymmetry. However, from the
perspective of quantization the ten--dimensional supergravities look
not so promising since they are non--renormalizable as field
theories containing gravitation usually are. At the same time
(Super)String Theory is believed to be a renormalizable and even
finite quantum theory in the ultraviolet limit, and therefore it
consistently describes quantum gravity. A field theoretical reason
behind this consistent quantum behavior is the contribution to
quantum corrections of an infinite tower of massive higher spin
excitations of the string, whose mass squared is proportional to
string tension and spin (e.g. in open string theory  $M_s^2\sim
T\,(s-1)\sim {1\over\alpha'}\,(s-1)$, where $s$ is the maximum spin
value of a state). Therefore, a better understanding of the dynamics
of higher spin states is important for the analysis of quantum
properties of String Theory.

Until recently the field of higher spins has remained a virgin
land cultivated by only a few enthusiasts. But higher spin field
theory may become a fashionable topic if a breakthrough happens in
understanding its basic problems.

Our experience in quantum field theory teaches us that massive
fields with spin 1 and higher are non--renormalizable unless their
mass was generated as a result of spontaneous breaking of a gauge
symmetry associated with corresponding massless gauge fields. So
first of all we should understand the structure of the theory of
massless higher spin fields.

In String Theory higher spin excitations become massless in the
limit of zero string tension. Thus in this limit one should
observe an enhancement of String Theory symmetry by that of the
massless higher spin fields, and one can regard string tension
generation as a mechanism of spontaneous breaking of the higher
spin symmetry. If the conjecture that String Theory is a
spontaneously broken phase of an underlying gauge theory of higher
spin fields is realized, it can be useful for better understanding
of string/M theory and of the (A)dS/CFT correspondence (see e.g.
\cite{V,BO,DNW,ss,dw0, bianchi} and references therein). This is
one of the motivations of the development of the theory of
interacting higher spin fields.

A direct but, perhaps, too involved way of studying the higher
spin string states would be the one in the framework of String
Field Theory, which itself is still under construction as far as
supersymmetric and closed strings are concerned. Another possibile
way is, as in the case of lower spin excitations, to derive an
effective field theory of higher spins and to study its properties
using conventional field theoretical methods.

In fact, higher spin field theory, both for massive and massless
fields, has been developed quite independently of String Theory
for a long period of time starting from papers by Dirac
\cite{dirac}, Wigner \cite{wigner}, Fierz and Pauli
\cite{fierz&pauli}, Rarita and Schwinger \cite{rarita&schwinger},
Bargmann and Wigner \cite{bargmann&wigner}, Fronsdal
\cite{fronsdal58}, Weinberg \cite{sw} and others. In last decades
a particular attention has been paid to massless higher spin
fields whose study revealed a profound and rich geometrical and
group--theoretical (conformal) structure underlying their
dynamics.

Understanding the interactions of higher spin fields is a main long
standing problem of the construction of the higher spin field
theory. The interaction problem already reveals itself when one
tries to couple higher spin fields to an electromagnetic field
\cite{fierz&pauli,singh&hagen} or to gravity
\cite{aragone&deser,berends&vanholten}, or to construct
(three--vertex) self--interactions
\cite{deserG,bengtsson&al,Berends&al}. In the case of massless
higher spin fields the problem is in introducing interactions in
such a way that they do not break (but may only properly modify)
gauge symmetries of the free higher spin field theory. Otherwise the
number of degrees of freedom in the interacting theory would differ
from that of the free theory, which apparently would result in
inconsistencies.

One should also note that the general (Coleman--Mandula and
Haag--Lopuszanski--Sohnius) theorem of the possible symmetries of
the unitary S--matrix of the quantum field theory in $D=4$
Minkowski space \cite{matrix} does not allow conserved currents
associated with symmetries of fields with spin greater than two to
contribute to the S--matrix. This no--go theorem might be overcome
if the higher spin symmetries would be spontaneously broken, as
probably happens in String Theory.

Another way out is that one should construct the interacting
higher spin field theory in a vacuum background with a non--zero
cosmological constant, such as the Anti de Sitter space, in which
case the S--matrix theorem does not apply. This has been realized
in \cite{fradkin&vasiliev}, where consistent interactions of
massless higher spin fields with gravity were constructed in the
first non--trivial (cubic) order. Until now the extension of these
results to higher orders in the coupling constant at the level of
the action has encountered difficulties of a group--theoretical
and technical nature related to the problem of finding the full
algebraic structure of interacting higher spin symmetries. As has
been noted in
\cite{bengtsson&al,Berends&al,fradkin&vasiliev,deserD}, such an
algebraic structure and consistent interactions should involve
higher derivative terms and infinite tower of fields with
increasing spins, and this again resembles the situation which we
have in String Theory. At the level of so called unfolded
equations of motion non--linear gauge field  models of interacting
massless higher spin fields have been constructed in
\cite{vas,progress,arbitraryD}.

To study the relation of higher spin field theory to superstring
theory one should work in ten--dimensional space--time. Here we
encounter a ``technical'' problem. In $D=4$ all states of higher
spin can be described either by the higher rank symmetric tensors
or spin tensors, since all tensor fields with mixed, symmetric and
antisymmetric, components can be related via Poincar\'e duality to
the symmetric tensors. This is not the case, for instance, in
$D=10$ where mixed symmetry tensor fields describe independent
higher spin modes and should be studied separately \cite{mixed}.
{}From the group--theoretical point of view this is related to the
fact that in $D=4$ the compact subgroup of the Wigner little
group, which is used to classify all the massless irreducible
representations of the Poincar\'e group, is $SO(2)$ whose Young
tableaux are  single symmetric rows \footnote{In the case of the
massive higher spin fields in $D=4$ the Wigner little group is
$SU(2)\sim SO(3)$, whose irreducible representations are also
described by only single row Young tableaux because of the
degeneracy of the antisymmetric three--dimensional matrix.}, while
in $D=10$ the compact subgroup of the little group is $SO(8)$
whose representations are described by Young tableaux with both
(symmetric) rows and (antisymmetric) columns. An essential
progress in studying the mixed symmetry fields has been made only
quite recently \cite{bpt}--\cite{alka}.

In these lectures, with the purpose of simplifying a bit the
comprehension of the material, we shall mainly deal with higher
spin fields described by symmetric tensors and spin tensors. The
article is organized according to its Contents.

These notes are not a comprehensive review but rather an attempt
to write an elementary introduction to only few aspects of higher
spin field theory and its history. I apologize to the authors
whose work has not been reflected in what follows.

\section{Free higher spin field theory}
\subsection{The choice of Lorentz representations for describing
higher spin fields. Symmetric tensors and spin--tensors}\label{FF}
 One of the possible choices is to associate potentials of integer higher
spin fields with symmetric tensors. In $D=4$ the symmetric tensors
describe all possible higher spin representations of the
Poincar\'e group because the antisymmetric second rank potentials
are dual to scalar fields and the three and four form potentials
do not carry physical degrees of freedom. As we have already
mentioned, this can also be understood using the fact that the
compact subgroup of the little group of the $D=4$ Lorentz group is
one--dimensional $SO(2)$. In higher dimensions, for instance in
$D=10$, the symmetric tensors do not embrace all the integer
higher spins, and one should also consider tensors with the
indices of mixed symmetry (symmetric and antisymmetric).

We shall restrict ourselves to the consideration of the symmetric
tensor fields $\phi_{m_1\cdots m_{s}}(x)$ which, under some
conditions to be discussed below, describe higher spin states of
an integer spin $s$. To describe the physical states of half
integer spin $s$ one should consider spinor tensor fields
$\psi^{\alpha}_{m_1\cdots m_{s-{1\over 2}}}(x)$
\footnote{This formulation of the higher spin fields is also called
the metric--like formulation because it is constructed as a
generalization of the metric $\phi_{mn}=g_{mn}$ formulation of
General Relativity. For non--linear extensions of higher spin field
theories another  formalism has proved to be useful. It is based on
a generalization of the description of gravity in terms the vielbein
and spin connection, and is called the frame--like formulation (see
\cite{frame} for a review and references).}.

In string theory symmetric tensor fields arise, for examples, as
string states obtained by acting on the vacuum by a single string
oscillator $a^{-n}_m$ with fixed integer $n$ (e.g. n=1)
$$
\phi_{m_1\cdots m_{s}}=a^{-1}_{m_1}\cdots a^{-1}_{m_s}|0>.
$$

Alternatively, the field strengths of half integer and integer
spin can be described by symmetric spin--tensors
$\varphi_{\alpha_1\cdots\alpha_{2s}}(x)$ depending on whether $s$
is half integer or integer \cite{sw}. The advantage of this
formulation is that all the spins (integer and half integer) are
treated on an equal footing.

In the Green--Schwarz formulation of the superstring such fields
arise as string states obtained by acting on the string vacuum
with an antisymmetrized product of different fermionic oscillators
$\theta^{-n}_\alpha$
$$\varphi_{\alpha_1\cdots\alpha_{2s}}=\theta^{[-1}_{\alpha_1}\cdots
\theta^{-2s]}_{\alpha_{2s}}|0>,
$$
where $n=1,\cdots 2s$ labels fermionic oscillator modes.

 We shall briefly consider the spin--tensor formulation in Subsection 3.4.

\subsection{Symmetric tensor description of massless higher spin
fields}

As it has already been mentioned, in quantum field theory  massive
fields with spin 1 and higher are not renormalizable unless their
mass is generated as a result of spontaneous breaking of a gauge
symmetry associated with corresponding massless gauge fields. So
first of all we should understand the structure of the theory of
massless higher spin fields and I shall concentrate on this
problem. The theory of massive higher spin fields and their
interactions (in particular with electromagnetic fields and
gravity) was discussed e.g. in
\cite{wigner,fierz&pauli,rarita&schwinger,fronsdal58,sw,chang,singh&hagen,ady,yura,dw,massive}
 .

Note that in $D=4$ the physical fields of spin $s\leq 2$ are part
of the family of the symmetric (spin) tensors. Their well known
equations of motion and gauge transformations are reproduced below
in a form suitable for the generalization to the case of the
higher spin fields
\begin{description}
\item[s=0~~] $\phi(x)$ --  scalar field,
$\partial_m\partial^m\phi\equiv \partial^2\,\phi=0$,  matter
field,
 no gauge symmetry;
\item[$\bf s={1\over 2}$] $\psi^\alpha(x)$ -- spinor field,
$\gamma^{m\alpha}_{~~~\beta}\partial_m\psi^\beta=\slash\!\!\!\!\!(\partial\psi)^\alpha=0$,
 matter field, no gauge symmetry;
\item[s=1~~] $\phi_m(x)=A_m(x)$ -- Maxwell field,
$\partial^mF_{mn}=\partial^2\, A_n-\partial_n\partial_m A^m=0$,
$\delta A_m=\partial_m\xi(x)$;
\item[$\bf s={3\over 2}$] $\psi^\alpha_m(x)$ --  Rarita--Schwinger field,
$\gamma_{mnp}\partial^n\psi^p=\slash\!\!\!\partial\psi_m-\partial_m\gamma^n\psi_n=0,
~~\delta\psi_m^\alpha=\partial_m\xi^\alpha(x) $;
\item[s=2~~] $\phi_{m_1m_2}(x)=g_{m_1m_2}(x)$ -- graviton,
$R_{m_1m_2}=0, ~~~\delta
g_{m_1m_2}=D_{m_1}\xi_{m_2}+D_{m_2}\xi_{m_1}$,\\
where $D_m=\partial_m+\Gamma_{mn}^{~~p}$ is the covariant
derivative and $\Gamma_{mn,p}={1\over
2}(\partial_pg_{mn}-\partial_mg_{np}-\partial_ng_{mp})$ is the
Christoffel connection;\\
in the linearized limit where the deviation of $g_{m_1m_2}(x)$
from the Minkowski metric $\eta_{m_1m_2}$ is infinitesimal the
Einstein equation and the diffeomorphisms reduce to\\
 $\partial^2\, g_{m_1m_2}-\partial_{m_1}\partial_n g^n_{~m_2}
-\partial_{m_2}\partial_n
g^n_{~m_1}+\partial_{m_1}\partial_{m_2}g^n_{~n}=0$, $\delta
g_{m_1m_2}=\partial_{m_1}\xi_{m_2}+\partial_{m_2}\xi_{m_1}$.
\end{description}

 Except for the scalar and the spinor field, all other
massless fields are gauge fields. The associated gauge symmetry
eliminates (unphysical) lower spin components of these fields and
thus ensures that they have a positive norm. So it is natural to
assume that all massless higher spin fields are also the gauge
fields with the gauge transformations being an appropriate
generalization of those of the Maxwell, Rarita--Schwinger and
Einstein field. In the linear (free field) approximation the
higher spin gauge transformations (for the integer and half
integer spins) have the form
\begin{eqnarray}\label{hsg}
&\delta\phi_{m_1\cdots m_{s}}(x)=\partial_{m_1}\xi_{m_2\cdots
m_{s}}+\partial_{m_2}\xi_{m_1\cdots m_{s}}+\cdots\equiv\sum
\partial_{m_1}\xi_{m_2\cdots m_{s}}\,, \quad\nonumber\\
&\delta\psi^\alpha_{m_1\cdots m_{s-{1\over 2}}}(x)=\sum
\partial_{m_1}\xi^\alpha_{m_2\cdots m_{s-{1\over 2}}}\,,
\end{eqnarray}
where $\sum$ will denote (almost everywhere) the symmetrized sum
with respect to all non-contracted vector indices.

\subsubsection{Free equations of motion}

We assume that the free equations of motion of the higher spin
fields are second order linear differential equations in the case of
the integer spins and the first order differential equations in the
case of the half integer spins. This is required by the unitary and
ensures that the fields have a positive--definite norm. The massless
higher spin equations have been derived from the massive higher spin
equations \cite{singh&hagen} by Fronsdal for bosons
\cite{fronsdal78} and by Fang and Fronsdal for fermions
\cite{fronsdal78a}, and studied in more detail in
\cite{deWit&Freedman}.

The  bosonic equations, which I shall denote by $G_{m_1\cdots
m_{s}}(x)$ are a natural generalization of the Klein-Gordon,
Maxwell and linearized Einstein equations
\begin{equation}\label{be}
G_{m_1\cdots m_{s}}(x)\equiv\partial^2\,\phi_{m_1\cdots
m_{s}}(x)-\sum
\partial_{m_1}\partial_n\phi^n_{~m_2\cdots
m_{s}}(x)+\sum \partial_{m_1}\partial_{m_2}\phi^n_{~nm_3\cdots
m_{s}}(x)=0\,.
\end{equation}

 The first order fermionic equations are a natural
generalization of the Dirac and Rarita--Schwinger equation
\begin{equation}\label{fe}
G^\alpha_{m_1\cdots m_{s-{1\over 2}}}(x)\equiv
(\slash\!\!\!\partial\psi)^{\alpha}_{m_1\cdots m_{s-{1\over
2}}}-\sum \partial_{m_1}(\gamma^n\psi)^{\alpha }_{nm_2\cdots
m_{s-{1\over 2}}}=0\,.
\end{equation}

\subsubsection{Constraints on higher spin symmetry
parameters and on higher spin fields}

We should now verify that the equations of motion (\ref{be}) and
(\ref{fe}) are invariant under gauge transformations (\ref{hsg}).
The direct computations give
\begin{equation}\label{dG}
\delta G_{m_1\cdots m_s}=3\sum
\partial^3_{m_1m_2m_3}\xi^n_{~nm_4\cdots m_s}, \quad \delta
G^\alpha_{m_1\cdots m_{s-{1\over 2}}}=-2\sum
\partial^2_{m_1m_2}\gamma^{n\alpha}_{~~~\beta}\,
\xi^\beta_{~nm_3\cdots m_{s-{1\over 2}}}\,,
\end{equation}
where $\partial^2_{m_1m_2}=\partial_{m_1}\partial_{m_2}$ and
$\partial^3_{m_1m_2m_3}=\partial_{m_1}\partial_{m_2}\partial_{m_3}$.

 We see that these variations vanish if  the parameters of
the transformations of the bosonic higher spin fields (for $s\geq
3$) are traceless
\begin{equation}\label{tl}
\xi^n_{~nm_4\cdots m_s}=0
\end{equation}
and the parameters of the transformations of the fermionic higher
spin fields for ($s\geq 5/2$) are $\gamma$--traceless
\begin{equation}\label{gtl}
(\gamma^{n}\xi)^\alpha_{~nm_3\cdots m_{s-{1\over 2}}}=0\,.
\end{equation}

 Other constraints in the theory of higher spins appear for
bosons with $s\geq 4$ and for fermions with $s\geq {7\over 2}$.
Since the theory is gauge invariant there should exist Bianchi
identities analogous to those in Maxwell and Einstein theory which
are identically satisfied. The Bianchi identities (or, equivalently,
integrability conditions) imply that the traceless divergence of the
left--hand--side of equations of motion must vanish identically.
This also implies that the currents of the matter fields if coupled
to the gauge fields are conserved.

For instance in Maxwell theory we have
\begin{equation}\label{mbi}
\partial_n(\partial_mF^{mn})\equiv 0,
\end{equation}
and, hence the electric current which enters the r.h.s. of the
Maxwell equations $\partial_mF^{mn}=J^m$ is conserved
$\partial_mJ^m=0$.

In the theory of gravity coupled to matter fields and described by
the Einstein equation
$$
R_{mn}-{1\over 2}g_{mn}R=T_{mn}
$$
 the energy--momentum conservation
$D_mT^{mn}=0$ is related to the Bianchi identity
\begin{equation}\label{ebi}
D_mR^m_{~n}-{1\over 2}D_nR^m_{~m}\equiv 0\,.
\end{equation}

The linearized form of (\ref{ebi}) generalized to the case of the
bosonic higher spin fields results in the following Bianchi
identity (or the integrability condition)
\begin{equation}\label{hsbi}
\partial_nG^n_{~m_2\cdots m_s}-{1\over
2}\sum \partial_{m_2}G^n_{~nm_3\cdots m_s}=-{3\over 2}\sum
\partial^3_{m_2m_3m_4}\phi^{np}_{~~npm_5\cdots m_s}\,,
\end{equation}
and in the case of the fermionic fields we have
\begin{equation}\label{fbi}
\partial_nG^{\alpha n}_{~~m_2\cdots m_{s-{1\over 2}}}-{1\over
2}\sum \partial_{m_2}G^{\alpha n}_{~~nm_3\cdots m_{s-{1\over
2}}}-{1\over 2}(\slash\!\!\!\partial \gamma^nG)^\alpha_{nm_2\cdots
m_{s-{1\over 2}}}=\sum \partial^2_{m_2m_3}(\gamma^n\psi)^{\alpha
p}_{~~npm_4\cdots m_{s-{1\over 2}}}\,.
\end{equation}
We see that the right--hand--sides of (\ref{hsbi}) and (\ref{fbi})
do not vanish identically and require that the bosonic fields with
$s\geq 4$ are double--traceless
\begin{equation}\label{dtr}
\phi^{np}_{~~npm_5\cdots m_s}=0
\end{equation}
and the fermionic fields with $s\geq {7\over 2}$ are
triple--gamma--traceless
\begin{equation}\label{trgtr}
(\gamma^n\gamma^p\gamma^r\psi)^{\alpha }_{~~nprm_4\cdots
m_{s-{1\over 2}}}\equiv (\gamma^n\psi)^{\alpha p}_{~~npm_4\cdots
m_{s-{1\over 2}}}=0\,.
\end{equation}
It turns out that for the consistency of the theory the fields
should satisfy the double--triple traceless conditions identically,
i.e. {\it off the mass shell}. Note that the double-- and
triple--traceless conditions are the strongest possible gauge
invariant algebraic constraints on the fields, provided that the
gauge parameters are traceless.

Physically the requirement of the double tracelessness, together
with the gauge fixing of higher spin symmetry, ensures that the
lower spin components contained in the symmetric tensor fields are
eliminated, so that only the massless states with helicities $\pm
s$ propagate. And as we know very well for lower spin fields
${1\over 2}\leq s\leq 2$, in $D=4$ each massless field has only
two physical degrees of freedom which are characterized by the
helicities $\pm s$.

Pure gauge degrees of freedom of the integer higher spin fields
can be eliminated by imposing gauge fixing conditions analogous to
the Lorentz gauge of the vector field and the de Donder gauge in
the case of gravity
\begin{equation}\label{dedonder}
\partial_p\phi^p_{~m_2\cdots m_s}-{1\over 2}\sum
\partial_{m_2}\phi^p_{~pm_3\cdots m_s}=0\,.
\end{equation}
Then the higher spin equations of motion (\ref{be}) reduce to the
Klein--Gordon equation $\partial^2\, \phi_{m_1\cdots m_s}=0$,
which implies that we indeed deal with {\it massless} fields.

Covariant gauge fixing condition for fermion fields are
\cite{deWit&Freedman}
\begin{equation}\label{gff}
\gamma^n\,\psi_{nm_2\cdots m_{s-{1\over 2}}}=0, \quad \Rightarrow
\quad \psi^n_{~nm_3\cdots m_{s-{1\over 2}}}=0\,.
\end{equation}
They reduce the field equation (\ref{fe}) down to the massless
Dirac equation.

  Thus, the double--and triple--traceless
constraints along with the gauge fixing conditions single out
physical components of the massless higher spin fields. Another
role of the double-- and triple--traceless constraints
(\ref{dtr}), (\ref{trgtr}) is that only when the fields
identically satisfy (\ref{dtr}) and (\ref{trgtr}), the higher spin
field equations (\ref{be}) and (\ref{fe}) can be obtained from
appropriate actions \cite{fronsdal78,deWit&Freedman}.

Unconstrained formulations of higher spin field dynamics will be
considered in the next Section.

\subsubsection{The free higher spin field actions}

For the bosonic fields the action in $D$--dimensional space--time
is
\begin{equation}\label{ba}
S_B=\int d^D x \left({1\over 2} \phi^{m_1\cdots
m_{s}}\,G_{m_1\cdots m_{s}}-{1\over
8}\,s(s-1)\,\phi_n^{~nm_3\cdots m_{s}}\,G^p_{~pm_3\cdots m_{s}}
\right)
\end{equation}
and for fermions
\begin{eqnarray}\label{fa}
S_F=&\int d^D x \left(-{1\over 2}\bar\psi^{m_1\cdots m_{s-{1\over
2}}}G_{m_1\cdots m_{s-{1\over 2}}}+{1\over
4}\,s\,\bar\psi^{m_2\cdots m_{s-{1\over
2}}n}\gamma_n\gamma^pG_{pm_2\cdots m_{s-{1\over
2}}}\right.\nonumber\\
 &\left.+{1\over 8}s(s-1)\bar\psi_n^{~nm_3\cdots m_{s-{1\over
2}}}G^p_{~pm_3\cdots m_{s-{1\over 2}}}\right)\,,
\end{eqnarray}
where $G_{m_1\cdots m_{s}}$ and $G^\alpha_{m_1\cdots m_{s-{1\over
2}}}$ stand for the left hand sides of the equations (\ref{be})
and (\ref{fe}).

The actions are invariant under the gauge transformations
(\ref{hsg}) with the traceless parameters (\ref{tl}) and
(\ref{gtl}), and the higher spin fields are supposed to be double
or gamma--triple traceless (\ref{dtr}), (\ref{trgtr}).

\section{Geometric aspects of free higher spin field theory}

The presence of the constraints on the gauge parameters and higher
spin fields in the formulation of Fronsdal and of Fang and
Fronsdal may look as an odd feature of the theory and point out
that such a formulation is incomplete. A modification of the
equations of motion (\ref{be}) and (\ref{fe}) which would remove
the constraints (\ref{tl}) and (\ref{gtl}) on the gauge parameters
and the double--traceless conditions (\ref{dtr}), (\ref{trgtr})
can be achieved in three different though related ways.

One of the ways to remove the tracelessness constraints is to use,
in addition to the physical higher spin field, an appropriate number
of auxiliary tensorial fields satisfying certain equations of
motion, as was shown in \cite{bpt} \footnote{In the case of massive
higher spin fields, auxiliary fields to construct higher spin field
actions were introduced by Fierz and Pauli \cite{fierz&pauli}.}. The
higher spin field equations remain lagrangian \cite{bpt,bkt}.

Another way was proposed by Francia and Sagnotti \cite{fs}. Its
key point is to renounce locality of the theory. It was shown that
the equations of motion of the unconstrained higher spin fields
and corresponding actions can be made invariant under the
unconstrained gauge transformations if they are enlarged with {\it
non--local} terms.
 A motivation of Francia and Sagnotti for removing constraints on
gauge parameters has been based on the observation that symmetries
of String Field Theory do not have such restrictions. Another
motivation was to find a more conventional {\it geometric} form of
the higher--spin field equations in terms of conditions on
generalized curvatures introduced in \cite{deWit&Freedman}. This
would be a generalization of the Maxwell and Einstein equations
written in terms of $F_{mn}$ and $R_{mn}$, respectively. The
choice of non--local terms in the higher spin field equations is
not unique. Choosing a suitable form of non--local equations one
manages to keep their Lagrangian nature. We will not go into
details of this formulation and address the interested reader to
\cite{fs,Bouatta}.

A third possibility of removing the constraints is to allow the
higher spin field {\it potentials} to satisfy {\it higher order}
differential equations, which can be constructed in a manifestly
gauge invariant way as conditions imposed on the higher spin field
curvatures. Let us consider this geometric formulation of the free
higher spin field theory in more detail. Actually, in $D=4$
space--time it has been constructed many years ago by Bargmann and
Wigner \cite{bargmann&wigner}. As we shall see, the higher order
derivative structure of the higher spin curvature equations does
not spoil the unitarity of the theory. These equations are
physically equivalent to the Fang--Fronsdal and Francia--Sagnotti
equations.

Generalized curvatures for the higher spin fields $\phi_{m_1\cdots
m_{s}}(x)$ and $\psi^\alpha_{m_1\cdots m_{s-{1\over 2}}}(x)$ which
are invariant under the {\it unconstrained} gauge transformations
(\ref{hsg}) can be constructed as a direct generalization of the
spin 1 Maxwell field strength
$$
F_{mn}=\partial_m\,A_n-\partial_n\,A_m
$$
and of the linearized Riemann tensor in the case of spin 2
\begin{equation}\label{Rim}
R_{m_1n_1,\,m_2n_2}=\partial_{m_1}\,\partial_{m_2}\,g_{n_1n_2}-\partial_{n_1}\,\partial_{m_2}\,g_{m_1n_2}
-\partial_{m_1}\,\partial_{n_2}\,g_{n_1m_2}+\partial_{n_1}\,\partial_{n_2}\,g_{m_1m_2}.
\end{equation}
Thus, for an arbitrary integer spin $s$ the gauge invariant
curvature is obtained by takin  $s$ derivatives of the field
potential $\phi_{n_1\cdots n_{s}}(x)$
\begin{eqnarray}\label{bis}
R_{m_1n_1,\,m_2n_2,\cdots,\,m_sn_s}=\partial_{m_1}\,\partial_{m_2}\cdots\,
\partial_{m_s}\,\phi_{n_1n_2\cdots\,n_s}
-\partial _{n_1}\,\partial_{m_2}\cdots\, \partial_{m_s}\,\phi_{m_1n_2\cdots\,n_s}\nonumber\\
\nonumber\\
 -\partial_{m_1}\,\partial_{n_2}\cdots\,
\partial_{m_s}\,\phi_{n_1m_2\cdots\,n_s}
+\partial_{n_1}\,\partial_{n_2}\cdots\,
\partial_{m_s}\,\phi_{m_1m_2\cdots n_s}+\cdots\nonumber\\
\nonumber\\ \equiv
\partial^s_{m_1\cdots\,m_s}\,\phi_{n_1\cdots\,n_s}-\sum\,(m_i\leftrightarrow n_i)\,.
\end{eqnarray}
Analogously, for an arbitrary half integer spin $s$ the curvature
is obtained by taking  $(s-{1\over 2})$ derivatives of the field
potential $\psi^\alpha_{n_1\cdots n_{s-{1\over 2}}}(x)$
\begin{eqnarray}\label{fis}
{\cal R}^\alpha_{m_1n_1,\,m_2n_2,\,\cdots\,,m_{s-{1\over
2}}n_{s-{1\over 2}} }=\partial_{m_1}\,\partial_{m_2}\cdots\,
 \partial_{m_{s-{1\over 2}}}\,\psi^\alpha_{n_1n_2\cdots\,n_{s-{1\over
2}}}-\partial_{n_1}\,\partial_{m_2}\cdots\,\partial_{m_{s-{1\over
2}}}\,\psi^\alpha_{m_1n_2\cdots\,
n_{s-{1\over 2}}}\nonumber\\
 \nonumber\\
 -\partial_{m_1}\,\partial_{n_2}\cdots \,\partial_{m_{s-{1\over
2}}}\,\psi^\alpha_{n_1m_2\cdots\,n_{s-{1\over
2}}}+\partial_{n_1}\,\partial_{n_2}\cdots \,\partial_{m_{s-{1\over
2}}}\,\psi^\alpha_{m_1m_2\cdots n_{s-{1\over 2}}}+\cdots\nonumber\\
\nonumber\\
 \equiv  \partial^{^{s-{1\over
2}}}_{m_1\cdots\,m_{s-{1\over
2}}}\,\psi^\alpha_{n_1\cdots\,n_{s-{1\over
2}}}-\sum\,(m_i\leftrightarrow n_i) \,.\hspace{30pt}~
\end{eqnarray}
In the right hand side of (\ref{bis}) and (\ref{fis}) it is
implied that the sum is taken over all the terms in which the
indices within the pairs of $[m_i,\,n_i]$ with the same label
$i=1,\cdots, \,s$ are antisymmetrized.

By construction similar to the Riemann tensor (\ref{Rim}), the
higher spin curvatures (\ref{bis}) and (\ref{fis})   are
completely symmetric under the exchange of any two pairs of their
antisymmetric indices and they obey for any pair of the
antisymmetric indices $[m_i\,n_i]$ the same Bianchi identities as
the Riemann tensor, {\it e.g.} for the bosonic spin $s$ field
\begin{equation}\label{symmetry}
R_{m_1n_1,\,m_2n_2,\cdots,\,m_sn_s}=-R_{n_1m_1,\,m_2n_2,\cdots,\,m_sn_s}
=R_{m_2n_2\,,m_1n_1,\cdots,\,m_sn_s}\,,
\end{equation}
\begin{equation}\label{B1}
R_{[m_1n_1,\,m_2]n_2,\cdots,\,m_sn_s}=0\,,
\end{equation}
\begin{equation}\label{B2}
\partial_{[l_1}R_{m_1n_1],\,m_2n_2,\cdots,\,m_sn_s}=0\,.
\end{equation}
On the other hand, if a rank $2[s]$  (spinor)--tensor (where $[s]$
is the integer part of $s$) possesses the properties
(\ref{symmetry})--(\ref{B2}), in virtue of the generalized
Poincar\'e lemma of \cite{Olver:1983,DuboisV,BB1} this tensor can
be expressed as an `antisymmetrized' $[s]$--th derivative of a
symmetric rank $[s]$ field potential, as in eqs. (\ref{bis}) and
(\ref{fis}).

Let us note that de Wit and Freedman \cite{deWit&Freedman}
constructed curvature tensors  out of the $[s]$ derivatives of the
symmetric higher spin gauge fields $\phi_{m_1\cdots\,m_s}(x)$ and
$\psi^\alpha_{m_1\cdots\, m_{s-{1\over 2}}}(x)$ in an alternative
way. Their curvatures have two pairs of the groups of $[s]$
symmetric indices and they are symmetric or antisymmetric under
the exchange of these groups of indices depending on whether $[s]$
is even or odd
\begin{equation}\label{dwf}
{\tilde R}_{m_1\cdots m_{[s]},\,n_1\cdots
n_{[s]}}=(-1)^{[s]}\,{\tilde R}_{n_1\cdots n_{[s]},\,m_1\cdots
m_{[s]}}\,.
\end{equation}
The de Wit--Freedman curvatures satisfy the cyclic identity, which
is a symmetric analog of (\ref{B1}),
\begin{equation}\label{cy}
{\tilde R}_{m_1\cdots m_{[s]},\,n_1\cdots
n_{[s]}}+\sum_{n_i}\,{\tilde R}_{n_1m_2\cdots
m_{[s]},\,m_1n_2\cdots n_{[s]}}=0\,,
\end{equation}
where $\sum_{n_i}$ denotes the symmetrized sum with respect to the
indices $n_i$. The form of the analogs of the differential Bianchi
identities (\ref{B2}) for ${\tilde R}_{m_1\cdots
m_{[s]},\,n_1\cdots n_{[s]}}$ is less transparent. They can be
obtained from (\ref{B2}) using the fact that the de Wit--Freedman
tensors (\ref{dwf}) are related to the generalized `Riemann'
tensors (\ref{bis}) and (\ref{fis}) by the antisymmetrization in
the former of the indices of each pair $[m_i,\,n_i]$. In what
follows we shall work with the generalized Riemann curvatures.

To produce the dynamical equations of motion of the higher spin
fields one should now impose additional conditions on the higher
spin curvatures. Since the curvatures (\ref{bis}) and (\ref{fis})
have the same properties (\ref{symmetry})--(\ref{B2}) as the
Riemann tensor (\ref{Rim}), and the linearized Einstein equation
amounts to putting to zero the trace of the Riemann tensor
\begin{equation}\label{trRim}
R^m{}_{n_1,mn_2}=R_{n_1n_2}=\partial^2\,\partial^2\,
g_{n_1n_2}-\partial_{n_1}\partial_m g^m_{~n_2}
-\partial_{n_2}\partial_m
g^m_{~n_1}+\partial_{n_1}\partial_{n_2}g^m_{~m}=0\,,
\end{equation}
one can naturally assume that in the case of the integer spins the
equations of motion are also obtained by requiring that the trace
of the generalized Riemann tensor (\ref{bis}) with respect to any
pair of indices is zero, e.g.
\begin{equation}\label{trbis}
R^m{}_{n_1,mn_2,m_3n_3,\cdots,\,m_sn_s}=0\,.
\end{equation}
In the case of the half integer spins we can assume that the
fermionic equations of motion are a generalization of the Dirac
and Rarita--Schwinger equations and that they are obtained by
putting to zero the gamma--trace of the fermionic curvature
(\ref{fis})
\begin{equation}\label{trfis}
(\gamma^{m_1}{\cal
R})^\alpha_{m_1n_1,\,m_2n_2,\,\cdots\,,m_{s-{1\over
2}}n_{s-{1\over 2}} }=0\,.
\end{equation}
The equations (\ref{trbis}) and (\ref{trfis}) are non--lagrangian
for $s>2$.

Let us remind the reader that, because of the gauge invariance of
the higher spin curvatures, eqs. (\ref{trbis}) and (\ref{trfis})
are invariant under the higher spin gauge transformations
(\ref{hsg}) with {\it unconstrained} parameters, and the higher
spin field potentials are also {\it unconstrained} in contrast to
the Fronsdal and Fang--Fronsdal formulation considered in the
previous Section \ref{FF}.

 The reason and
the price for this is that eqs. (\ref{trbis}) and (\ref{trfis})
are higher order differential equations, which might cause a
problem with unitarity of the quantum theory. However, as we shall
now show the equations (\ref{trbis}) and (\ref{trfis}) reduce to,
respectively, the second and first order differential equations
for the higher spin field potentials related to those of Fronsdal
(\ref{be}) and of Fang and Fronsdal (\ref{fe}).

\subsection{Integer spin fields}
In the integer spin case, analyzing the form of the left hand side
of eq. (\ref{trbis}) in terms of the gauge field potential
(\ref{bis}) one gets the higher spin generalization of the {spin~3}
Damour--Deser identity \cite{DD} which relates the trace of the
higher spin curvature to the left hand side of the Fronsdal
equations, namely
\begin{eqnarray}
tr \,R_{m_1n_1,\cdots,\,m_sn_s}
=R_{m_1n_1,m_2n_2,\cdots,\,n_{s-1}m,\,n_s}{}^{m}
~{\hspace{150pt}}\nonumber\\
~{\hspace{10pt}}=\partial^{s-2}_{m_1m_2\cdots \,m_{s-2}}\,
{G}_{n_1\cdots\,n_{s-2}n_{s-1}n_s}-\sum_{i=1}^{i=s-2}\,(m_i\leftrightarrow
n_i)\,\label{DD}
\end{eqnarray}
where the symmetric tensor $G(x)$ stands for the left hand side of
the Fronsdal equations (\ref{be}) (sometimes called the ``Fronsdal
kinetic operator"), and the indices $[m_i,n_i]$ with
$(i=1,\cdots,\,s-2)$ are anti--symmetrized.

When the curvature tensor satisfies the tracelessness condition
(\ref{trbis}) the left hand side of eq. (\ref{DD}) vanishes, which
implies that the tensor $G$ is $\partial^{s-2}$--closed. In virtue
of the generalized Poincar\'e lemma \cite{Olver:1983,DuboisV,BB1}
this means that (at least locally) $G$ is $\partial^3$--exact,
{\it i.e.} has the form \cite{BB2}
\begin{equation}
{G}_{n_1\cdots\,n_s}=\sum\,\partial^3_{n_1n_2n_3}\,
\rho_{n_4\cdots\,n_s}\,, \label{comp}
\end{equation}
where the sum implies the symmetrization of all the indices $n_i$
and $\rho(x)$ is a symmetric tensor field  of rank $(s-3)$ called
`compensator' since its gauge transformation
\begin{equation}\label{drho}
\delta\,\rho_{n_1\cdots\,n_{s-3}}=3\,\xi^{m}{}_{mn_1\cdots\,n_{s-3}}
\end{equation}
compensates the non--invariance (\ref{dG}) of the kinetic operator
$G(x)$ under the unconstrained local variations (\ref{hsg}) of the
gauge field potential $\phi(x)$. Eq. (\ref{comp}) was discussed in
\cite{fs,am}. Here we have obtained it from the geometric equation
on the higher spin field curvature (\ref{trbis}) following ref.
\cite{BB2} where such a derivation was carried out for a generic
mixed symmetry field \footnote{For a relevant earlier discussion
of the relationship of the equations on the Riemann and de
Wit--Freedman curvatures to the equations of motion of symmetric
(spinor)--tensor field potentials see \cite{collins}. }.

The trace of the gauge parameter (\ref{drho}) can be used to
eliminate the compensator field. Then eq. (\ref{comp}) reduces to
the Fronsdal equation (\ref{be}) which is invariant under residual
gauge transformations with traceless parameters.

\subsection{A non--local form of the higher spin equations} We
shall now demonstrate how the higher spin field equations with the
compensator (\ref{comp}) are related to the non--local equations
of \cite{fs}.  We shall consider the simple (standard) example of
a gauge field of spin 3. The case of a generic spin $s$ can be
treated in a similar but more tedious way. In a somewhat different
way the relation of the compensator equations (\ref{comp}) to
non--local higher spin equations was discussed in the second paper
of \cite{fs}.

For the {\it spin} 3 field the compensator equation takes the form
\begin{equation}\label{3}
G_{mnp}:=\partial^2\,\,\phi_{mnp}-3\partial_q\,\partial_{(m}\,\phi^{~~~q}_{np)}
+3\partial_{(m}\,\partial_n\,\phi_{p)q}^{~~~q}=\partial^3_{mnp}\,\rho(x)\,,
\end{equation}
where $()$ stand for the symmetrization of the indices with weight
one and $\rho(x)$ is the compensator, which is a scalar field in
the case of spin 3.

We now take the derivative and then the double trace of the left
and the right hand side of this equation and get
\begin{equation}\label{dg1}
\partial_m\,G^{mn}_{~~~n}=\partial^2\,\partial^2\,\rho(x)\,.
\end{equation}
Modulo the {\it doubly harmonic} zero modes $\rho_0(x)$,
satisfying $\partial^2\,\partial^2\,\rho_0(x)=0$, one can solve
eq. (\ref{dg1}) for $\rho(x)$ in a non--local form
\begin{equation}\label{rho}
\rho(x)={1\over
\partial^2\,\partial^2}\,\partial_m\,G^{mn}_{~~~n}\,.
\end{equation}
Substituting this solution into the spin 3 field equation
(\ref{3}) we get one of the non--local forms of the spin 3 field
equation constructed in \cite{fs}
\begin{equation}\label{FS}
G_{mnp}:=\partial^2\,\,\phi_{mnp}-3\partial_q\,\partial_{(m}\,\phi^{~~~q}_{np)}
+3\partial_{(m}\,\partial_n\,\phi_{p)q}^{~~~q}={1\over
\partial^2\,\partial^2}\,\partial^3_{mnp}\, (\partial_q\,G^{qr}_{~~~r})\,.
\end{equation}

Let us now consider a more complicated example of {\it spin} 4. In
the Fronsdal formulation, the fields of spin 4 and higher feature
one more restriction: they are double traceless. We shall show how
this constraint appears upon gauge fixing the compensator
equation, which for the spin 4 field has the form
\begin{equation}\label{4}
G_{mnpq}:=\partial^2\,\,\varphi_{mnpq}-4\partial_r\,\partial_{(m}\,\varphi^{~~~~r}_{npq)}
+6\partial_{(m}\,\partial_n\,\varphi_{pq)r}^{~~~~r}
=4\partial^3_{(mnp}\,\rho_{q)}(x)\,.
\end{equation}
Taking the double trace of (\ref{4}) we have
\begin{equation}\label{dtr1}
G^{mn}_{~~~mn}=3\partial^2\,\,\varphi^{mn}_{~~~mn}=4\partial^2\,\,\partial_m\,\rho^m\,.
\end{equation}
Taking the derivative of (\ref{4}) and the double trace we get
\begin{equation}\label{ddtr}
\partial_m\,G^{mn}_{~~~np}=\partial^2\,\partial^2\,\rho_p+3\partial_p\,\partial^2\,\,\partial_m\,\rho^m=
\partial^2\,\partial^2\rho_p+{3\over 4}\,\partial_p\,G^{mn}_{~~~mn}\,,
\end{equation}
where we have used (\ref{dtr1}) to arrive at the right hand side
of (\ref{ddtr}).

From (\ref{ddtr}) we find that, modulo the zero modes $\rho^p_0$
of $\partial^2\,\partial^2\,\rho^p_0=0$, the compensator field is
non--locally expressed in terms of the (double) trace of the
Fronsdal kinetic term
\begin{equation}\label{rhom}
\rho_p={1\over
\partial^2\,\partial^2}\,(\partial_m\,G^{mn}_{~~~np}-{3\over
4}\,\partial_p\,G^{mn}_{~~~mn})\,.
\end{equation}
Substituting (\ref{rhom}) into (\ref{4}) we get  one of the
non--local forms \cite{fs} of the spin 4 field equation \cite{fs}.

Consider now the following identity
\begin{equation}\label{bi}
\partial_q\,G^{q}_{~~mnp}-\,\partial_{(m}\,G^q_{~~np)q}=-{3\over
2}\partial_m\,\partial_n\,\partial_p\,\phi^{qr}_{~~~qr}=-{
2}\,\partial_m\,\partial_n\,\partial_p\,(\partial_q\,\rho^q)\,.
\end{equation}
On the other hand, from (\ref{dtr1}) and (\ref{bi}) it follows
that modulo a linear and quadratic term in $x^m$ (which can be put
to zero by requiring an appropriate asymptotic behaviour of the
wave function at infinity) the double trace of the gauge field
$\phi(x)$ is proportional to the divergence of $\rho_q(x)$
\begin{equation}\label{phirho}
\phi^{mn}{}_{mn}={4\over 3}\,\partial_q\,\rho^q\,.
\end{equation}
Therefore, when we partially fix the gauge symmetry by putting
$\rho_q(x)=0$, the double trace of the gauge field also vanishes
and we recover the Fronsdal formulation with the traceless gauge
parameter and the double traceless gauge field.

\subsection{Half integer spin fields}
 Let us generalize the previous consideration to the case of
fermions. The fermionic spin--$s$ field strength ${\cal R}^\alpha$
is the spinor--tensor
\begin{equation}\label{calRcurv}
{\cal R}^\alpha_{~m_1n_1,\cdots,\,m_{s-{1\over 2}}n_{s-{1\over
2}}}(x)\,.
\end{equation}
 It satisfies the Bianchi identities analogous to
 (\ref{B1}), (\ref{B2}) and thus can be expressed in terms of $s-{1\over 2}$ derivatives of
  a fermionic field potential (\ref{fis}).

The fermionic generalization of the Damour--Deser identity is
\begin{equation}
(\gamma^m\,{\cal R})^\alpha_{mn_1,m_2n_2,\cdots\,m_{s-{1\over
2}}n_{s-{1\over 2}}}=\partial^{s-{3\over
2}}_{m_1\cdots\,m_{s-{3\over 2}}}\,
{G^\alpha}_{n_1\cdots\,n_{s-{3\over 2}}n_{s-{1\over
2}}}\,-\sum_{i=1}^{i={s-{3\over 2}}}\, (m_i\leftrightarrow
n_i)\,,\label{DDgenf}
\end{equation}
where the some is taken over terms with the indices $[m_i,n_i]$
anti--symmetrized, and the Fang--Fronsdal fermionic kinetic
operator $G^\alpha$ acting on the gauge field $\psi^\alpha$ is
defined in (\ref{fe}). The field strength (\ref{fis}) is invariant
under the {\it unconstrained} gauge transformations (\ref{hsg}).

 When the fermionic
field strength satisfies the $\gamma$--tracelessness condition
(\ref{trfis})
$$
(\gamma^m\,{\cal R})^\alpha_{mn_1,m_2n_2,\cdots\,m_{s-{1\over
2}}n_{s-{1\over 2}}}=0\,,
$$
eq. (\ref{DDgenf}) implies that $G^\alpha$ is $\partial^{s-{3\over
2}}$--closed. Since $\partial^{s+{1\over 2}}\equiv 0$, by virtue
of the generalized Poincar\'e lemma $G^\alpha$ is
$\partial^2$--exact
\begin{equation}
{G^\alpha}_{n_1\cdots\,n_{s-{1\over 2}}}=\sum
\,\partial_{n_1}\,\partial_{n_2}\,
\rho^\alpha_{n_3\cdots\,n_{s-{1\over 2 }}}\,, \label{compf}
\end{equation}
where $\sum$ implies the symmetrization of all the indices $n_i$.

 Equation (\ref{compf}) is the
compensator equation given in \cite{fs,am}. The demonstration of
its relation to the gamma--traceless part of the fermionic higher
spin field strength has been given in \cite{D4610}.

The gauge variation of $G^\alpha(x)$ has been presented in
(\ref{dG}). It is compensated by a gauge shift of the field
$\rho^\alpha(x)$ given by the $\gamma$--trace of the gauge
parameter
\begin{equation}\label{rhogf}
\delta\rho^\alpha_{n_3\cdots\,n_{s-{1\over 2 }}}=-2\sum
\gamma^{m\alpha}_{~~~\beta}\, \xi^\beta_{~mn_3\cdots\,
n_{s-{1\over 2}}}\,.
\end{equation}
Thus, the compensator can be gauged away by choosing a gauge
parameter $\xi^\alpha(x)$ with the appropriate $\gamma$--trace.
Then, the equations of motion of the gauge field $\psi^\alpha(x)$
become the first order differential equations (\ref{fe}) which are
invariant under the gauge transformations (\ref{hsg}) with
$\gamma$--traceless parameters.

Alternatively, one can get non--local Francia--Sagnotti equations
for fermions by taking a particular non--local solution for the
compensator field in terms of the fermionic kinetic operator
$G^\alpha$. As a simple example consider the $s=5/2$ case. Eq.
(\ref{compf}) takes the form
\begin{equation}\label{s5/2}
G^\alpha_{mn}:={\not {\! \partial}
}\psi^\alpha_{mn}-2\partial_{(m}\,(\gamma^q\psi)^\alpha_{n)q}
=\partial_m\,\partial_n\,\rho^\alpha(x)\,.
\end{equation}
Taking the trace of (\ref{s5/2}) we get
\begin{equation}\label{frho}
\partial^2\,\,\rho^\alpha=G^{\alpha p}_{~~~p}\,.
\end{equation}
Hence, modulo the zero modes $\rho^\alpha_0(x)$ of the
Klein--Gordon operator $\partial^2\,\,\rho^\alpha_0=0$ the
compensator field is non--locally expressed in terms of the trace
of $G^\alpha_{mn}$
\begin{equation}\label{nfrho}
\rho^\alpha={1\over \partial^2\,}\,G^{\alpha p}_{~~~p}\,.
\end{equation}
Substituting (\ref{nfrho}) into (\ref{s5/2}) we get the
Francia--Sagnotti equation for the fermionic field of spin 5/2 in
the following form
\begin{equation}\label{FS5/2}
G^\alpha_{mn}:={\not {\! \partial}
}\psi^\alpha_{mn}-2\partial_{(m}\,(\gamma^q\psi)^\alpha_{n)q}
={1\over \partial^2\,}\,\partial_m\,\partial_n\,G^{\alpha
p}_{~~~p}\,.
\end{equation}

In the same way one can relate the compensator equations for an
arbitrary half integer spin field to the corresponding non--local
field equation. As in the bosonic case, one can find that for
$s\geq {7\over 2}$ the triple--gamma trace of the fermionic gauge
field potential is expressed in terms of the $\gamma$--trace and the
divergence of the compensator field and thus vanishes in the
'Fronsdal gauge' $\rho^\alpha=0$.

We have thus reviewed various formulations of free higher spin
field dynamics.

\section{The interaction problem}
As we have seen, the free theory of higher spin fields, both
massless and massive, exists and can be formulated in a
conventional field--theoretical fashion using the action
principle. An important problem, which still has not been
completely solved, is to introduce interactions of the higher spin
fields. Probably, String Field Theory should give the answer to
this problem if someone manages to extract the corresponding
information from the String Field Theory action. This itself is a
highly non--trivial problem which has not been realized yet.

So far the study of the problem of higher spin interactions has
been undertaken mainly in the framework of the standard
field--theoretical approach, and I would now like to review main
obstacles which one encounters in the way of constructing an
interacting {\sl massless} higher spin field theory.

One may consider self--interactions of fields of the same spin, or
interactions among fields of different spin. In the first case,
for example, the construction of consistent self--interactions of
massless vector fields results in either the non--abelian
Yang--Mills theory \cite{Berends&al} or in the non--linear
Dirac--Born--Infeld generalization of Maxwell theory. A consistent
way of introducing the self--interaction of the spin 2 field leads
to the Einstein theory of gravity \cite{deserG}. Consistency
basically means that the introduction of interactions should not
break the gauge symmetry, but may only modify it in a suitably
way.

An example of interactions among fields of different spin is the
universal gravitational interaction of the matter and gauge
fields. So, the construction of the theory of interactions of
higher spin fields with gravity is an important part of the
general interaction problem and it actually exhibits all aspects
of the general problem.

\subsection{Simple supergravity}

To see what kind of problems with the higher spin interactions
arise, let us first consider the example of coupling to
four--dimensional gravity the field of spin 3/2
\cite{fierz&pauli,rarita&schwinger} historically called the
Rarita--Schwinger field. This is an instructive example which
shows how the consistency of gravitational coupling leads to
supergravity \cite{vs,sg}, the theory invariant under local
supersymmetry in which the spin 3/2 field becomes the superpartner
of the graviton, called gravitino.

The general coordinate invariance of the complete non--linear
gravitational interactions requires that in the free field
equations partial derivatives get replaced with covariant
derivatives and the vector indices are contracted with the
gravitational metric $g_{mn}(x)$. So the free Rarita--Schwinger
equation
$$
\gamma^{mnp}\,\partial_n\psi_p=0
$$
should be generalized to include the interaction with gravity as
follows
\begin{equation}\label{rsg}
G^m=\gamma^{mnp}\, D_n\psi_p=0,
\end{equation}
where $\gamma_n=e_n^a(x)\gamma_a$ are the gamma--matrices
contracted with the vielbein $e_n^a(x)$\linebreak which is related
to the metric in the standard way $g_{mn}=e_m^ae_n^b\eta_{ab}$,
and
$D_m=\partial_m+\Gamma_{mn}^{~~~p}+\omega_{m\alpha}^{~~~\beta}$ is
the covariant derivative which contains the Christoffel symbol
$\Gamma_{mn}^{~~p}$ and a spin connection
$\omega_{m\alpha}^{~~~\beta}$ acting on spinor indices.

In the presence of the Rarita--Schwinger field the
right--hand--side of the Einstein equations acquires the
contribution of an energy--momentum tensor of the spin 3/2 field
\begin{equation}\label{eem}
R_{mn}-{1\over 2}g_{mn}R=T_{mn}(\psi)\,, \quad {\rm or}\quad
R_{mn}-T_{mn}+{1\over 2}\,g_{mn}\, T_l^{~l}=0\,.
\end{equation}
The explicit form of $T_{mn}$ is not known until a consistent
interacting theory is constructed.

Consider now the variation of the Rarita--Schwinger field equation
under the gauge transformations
\begin{equation}\label{fgtg}
\delta\psi^\alpha_m=D_m\xi^\alpha,
\end{equation}
which are the general covariant extension of the free field gauge
transformations. The variation of eq. (\ref{rsg}) is
\begin{equation}\label{DG}
\delta G^m=\gamma^{mnp}D_nD_p\xi={1\over
2}\gamma^{mnp}[D_n,D_p]\xi \sim
\gamma^{mnp}R_{np,qr}\gamma^{qr}\xi\sim R^m_{~~n}\gamma^n\xi\,,
\end{equation}
where the commutator of $D_n$ produces the Riemann curvature which
I schematically write as
\begin{equation}\label{R}
[D_m,D_n]\sim R_{mn,pq}\gamma^{pq}\,,
\end{equation}
and the last term in (\ref{DG}) is obtained by use of
$\gamma$--matrix identities.

We thus see that the gauge variation of the Rarita--Schwinger
equation is proportional to the Ricci tensor. It is zero if the
Ricci tensor is zero, i.e. when the gravitational field satisfies
the Einstein equations in the absence of the matter fields. This
is satisfactory if we are interested in the dynamics of the spin
3/2 field in the external background of a  gravitational field,
such as free gravitational waves, for example. But if we would
like to consider a closed graviton--spin 3/2 system, then the
Ricci tensor is non--zero, since the Einstein equations take the
form (\ref{eem}), and the variation (\ref{DG}) does not vanish. To
improve the situation we should require that the graviton also
non--trivially varies under the gauge transformations with the
spinorial parameter $\xi^\alpha(x)$ as follows
\begin{equation}\label{dg}
\delta g_{mn}={i\over 2}
(\bar\psi_n\gamma_m\xi+\bar\psi_m\gamma_n\xi),
\end{equation}
and we should take into account this variation of the graviton in
the Rarita--Schwinger equation. This will result in the following
variation of the Rarita--Schwinger equation
\begin{equation}\label{dGcom}
\delta G_m\sim (R_{mn}-T_{mn}+{1\over 2}\,g_{mn}\,
T_l^{~l})\gamma^n\xi\,,
\end{equation}
provided we also add appropriate second--order and fourth--order
fermionic terms into the definition of the covariant derivative
$D_m$, into the variation  of $\psi_m$ (\ref{fgtg}) and into the
definition of its energy--momentum tensor. We observe that the
variation of the Rarita--Schwinger equation has become
proportional to the Einstein equation and hence vanishes.

Thus, by modifying the gauge transformations of the
Rarita--Schwinger field, of the graviton and by appropriately
modifying the Rarita--Schwinger and the Einstein equations we have
achieved that under the gauge (supersymmetry) transformations with
the fermionic parameter $\xi^\alpha$ the Rarita--Schwinger and the
Einstein equations transform into each other and hence
consistently describe the coupling of the spin--3/2 field to
gravity.

What we have actually obtained is a simple $D=4$ supergravity
\cite{sg} which is invariant under local supersymmetry
transformations. It is amazing that supergravity was not discovered
much earlier than the 70s  by people studied the massless spin--3/2
field, using the above reasoning for the construction of a
consistent gravity -- spin--3/2 field interacting system. In this
respect let us cite what Fierz and Pauli (\cite{fierz&pauli}, page
226) wrote about the massless spin 3/2 field: ``Whereas the theory
for the spin value 2 has an important generalization for force
fields, namely the gravitational theory, we here [in the case of
spin 3/2] have no such a connection with a known theory. To get a
generalization of the theory with interactions one would first of
all have to find a physical interpretation of the gauge group, and
the conservation theorem connected with this group".

\subsection{Gravitational interaction of a spin $5\over 2$
field}

Let us now, by analogy with supergravity, try to couple to gravity
in four dimensions a field of spin ${5\over 2}$
\cite{aragone&deser} which is described by the spin--tensor field
$\psi^\alpha_{m_1m_2}$. Again, the general coordinate invariance
of the gravitational interactions requires that in the free field
equations partial derivatives get replaced with covariant
derivatives and the vector indices are contracted with the
gravitational metric $g_{mn}(x)=\eta_{mn}+\phi_{mn}(x)$, where
$\phi_{mn}(x)$ is the deviation of the metric from the flat
background which at the moment, for simplicity, we consider to be
small and satisfy free spin 2 equations of motion. Thus the
straightforward generalization of the equations of motion of the
 spin ${5\over 2}$ field which describes its ``minimal'' interaction with gravity is
\begin{equation}\label{5/2}
G_{\alpha
m_1m_2}=i\gamma^n_{\alpha\beta}(D_n\psi^{\beta}_{m_1m_{2}}
-D_{m_1}\psi^{\beta}_{nm_2} -D_{m_2}\psi^{\beta}_{nm_1})=0\,.
\end{equation}
In (\ref{5/2}) we have restored the imaginary unit $i$ for further
comparison with the massive Dirac equation
$(i\slash\!\!\!\!\partial-m)\psi=0$.

We should now check whether these equations are invariant under
the covariant modification of the higher spin gauge
transformations
\begin{equation}\label{chsg}
\delta\psi^\alpha_{m_1 m_2}(x)=
D_{m_1}\xi^\alpha_{m_2}+D_{m_2}\xi^\alpha_{m_1}\,.
\end{equation}
The gauge variation of the equations of motion (\ref{5/2}) is
\cite{deWit&Freedman}
\begin{equation}\label{vari}
\delta G_{m_1m_2}= i
[R_{m_1n}\gamma^n\xi_{m_2}+R_{m_2n}\gamma^n\xi_{m_1}
-(R_{nm_1m_2}^{~~~~~~~p}+R_{nm_2m_1}^{~~~~~~~p})\gamma^n\xi_p]\,,
\end{equation}
where I have suppressed the spinor index. We see that in a general
gravitational background the variation does not vanish because of
the presence of the Ricci tensor in the first two terms and of the
Riemann tensor in the last term. If the bare Riemann tensor did
not appear the spin ${5\over 2}$ field equations would at least
admit interactions with gravitational fields satisfying the free
Einstein equations $R_{mn}=0$. As we have discussed, this is the
case for the gravitino field of spin ${3\over 2}$ whose local
supersymmetry transformations produce only the Ricci tensor term
in the variation of the Rarita--Schwinger equations. Introducing
an appropriate supersymmetry variation of the graviton, one
insures that the variation of the Rarita--Schwinger equations is
proportional to the Einstein equations with the r.h.s. to be the
energy--momentum tensor of the gravitino field. However, in the
case of the higher spin fields with $s\geq {5\over 2}$ the bare
Riemann tensor always appears (as part of the Weyl tensor) in the
variation of the field equations, and no way has been found to
cancel such terms by adding non--minimal interaction terms and/or
modifying the higher spin symmetry transformations (including that
of the graviton), when the zero limit of the gravitation field
corresponds to the flat Minkowski space (i.e. when the
cosmological constant is zero).

So the conclusion has been made that in a space--time with zero
cosmological constant it is not possible to construct a consistent
gauge theory of interacting higher spin fields, which is in
agreement with the general theorem of the possible symmetries of
the S--matrix. But, as happens with many no--go theorems, sooner
or later people find a way to circumvent them. In the case of the
higher spins the way out has been found in constructing the theory
in the AdS space, which has a non--zero cosmological constant
$\Lambda$.

In the bosonic case Fronsdal and in the fermionic case Fang and
Fronsdal \cite{fronsdalAdS} have generalized the free higher--spin
field equations and actions to the AdS background. For instance,
the equation of motion of the spin ${5\over 2}$ field  takes the
following form
\begin{equation}\label{ads5/2}
G^\alpha_{m_1m_2}=i\gamma^{n\alpha}_{~~~\beta}(\nabla_n\psi^{\beta}_{m_1m_{2}}
-\nabla_{m_1}\psi^{\beta}_{nm_2}
-\nabla_{m_2}\psi^{\beta}_{nm_1})-2 \Lambda^{1\over
2}\psi^\alpha_{ m_1m_{2}}=0\,.
\end{equation}
One can notice that the last term in (\ref{ads5/2}) resembles a
mass term of the spin ${5\over 2}$ field, however the field has
the number of physical degrees of freedom equal to that of the
corresponding massless field in flat space, i.e. two states with
helicities $\pm {5\over 2}$. This is because the equation of
motion is invariant under the following gauge transformations
\begin{equation}\label{adsgt}
\delta\psi^\alpha_{m_1m_2}(x)= \nabla_{m_1}
\xi^\alpha_{m_2}+\nabla_{m_2}\xi^\alpha_{m_1}\,.
\end{equation}
In (\ref{ads5/2}) and (\ref{adsgt})
$\nabla_m=D_{m}+{i\Lambda^{1\over 2}\over 2} \gamma_{m}$ is a so
called $SO(2,3)$ covariant derivative, and $D_m$ is the standard
covariant derivative in the AdS space whose Riemann curvature has
the well known form
\begin{equation}\label{RAdS}
R_{~mnp}^{(AdS)~q}=-\Lambda(\delta^{~q}_mg_{np}-\delta^{~q}_ng_{mp})\,,
\end{equation}
Remember also that the AdS metric is conformally flat
$g^{AdS}_{mn}=(1-\Lambda x^px_p)^{-2}\eta_{mn}$.
 Note that the gamma--matrix $\gamma_n$ entering (\ref{ads5/2})
carries a curved vector index, it is hence non--constant and is
related to the constant Dirac matrix $\gamma_a$ (carrying a local
Lorentz index) via the vielbein $e_m^a(x)$ of the AdS space
$\gamma_n=e_m^a(x)\gamma_a$.

It is important to notice that when acting on a spinor field
$\psi^\alpha(x)$ the commutator of $\nabla_m$ is zero
\begin{equation}\label{comf}
[\nabla_m,\nabla_n]\psi^\alpha=0\,
\end{equation}
while acting on a vector field $V_p$ the commutator is
\begin{equation}\label{comv}
[\nabla_m,\nabla_n]V_p=[D_m,D_n]V_p=-R^{(AdS)~q}_{~mnp}V_q\,.
\end{equation}
For the spin-tensor fields $\psi^{\alpha}_{m_1\cdots m_{s-{1\over
2}}}$ we thus have
\begin{equation}\label{comfv}
[\nabla_n,\nabla_p]\psi^{\alpha}_{m_1\cdots m_{s-{1\over
2}}}=[D_m,D_n]\psi^{\alpha}_{m_1\cdots m_{s-{1\over 2}}}=-\sum
R^{(AdS)~q}_{~mnm_1}\psi^{\alpha}_{qm_2\cdots m_{s-{1\over 2}}}\,.
\end{equation}
Note also that $\nabla_m$ does not annihilate the gamma matrix
$\gamma_n$
\begin{equation}\label{ng}
\nabla_m\,\gamma_n={i\Lambda^{1\over 2}\over
2}[\gamma_m,\gamma_n]\,.
\end{equation}

Thus the gauge invariant field equations for higher spin fields in
AdS do exist. If we now consider fluctuations of the gravitational
field around the AdS background the gauge variation (\ref{vari})
of the equations (\ref{adsgt}) will again have contributions
similar to
 (\ref{ads5/2}) of the bare Riemann curvature of the
fluctuating gravitational field
\begin{equation}\label{variR}
\delta G_{m_1m_2} = -{i}
(R_{nm_1m_2}^{~~~~~~~p}+R_{nm_2m_1}^{~~~~~~~p})\gamma^n\xi_p+
\cdots \,,
\end{equation}
where $\cdots$ stand for harmless terms, which can be canceled by
an appropriate modification of the gauge transformations. As has
been first noticed by Fradkin and Vasiliev in 1987
\cite{fradkin&vasiliev}, because of the non--zero {\it
dimensionful} cosmological constant of the background, it is now
possible to modify the field equation (\ref{ads5/2}) such that the
variation of an appropriate additional term will cancel the
dangerous Riemann curvature term in (\ref{variR}), at least in the
first order of the perturbation of the gravitational field. For
the spin ${5\over 2}$ field the appropriate term describing its
non--minimal coupling to a gravitational fluctuation is
\begin{equation}\label{nonl}
\triangle G_{m_1m_2}={i\over
{2\Lambda}}(R_{pm_1m_2q}+R_{pm_2m_1q})\,{\slash\!\!\!\!\nabla}\psi^{pq}\,,
\end{equation}
where $R_{pm_1m_2q}$ is the Riemann curvature corresponding to the
deviation of the gravitational field from the AdS background. Such
a term can be obtained from a cubic interaction term in the spin
${5\over 2}$ action
\begin{equation}\label{s3}
S_{int}={i\over \Lambda}\int d^D x\,\sqrt{g}
\left\{\bar\psi^{m_1m_2}R_{pm_1m_2q}\,{\slash\!\!\!\!\nabla}\psi^{pq}+
\bar\psi^{m_1m_2}(\nabla_r\gamma^r
R_{pm_1m_2q}\,)\psi^{pq}\right\}\,.
\end{equation}
Note that the cosmological constant enters the interacting term
(\ref{nonl}), (\ref{s3}) in a non--polynomial way. Therefore, such
terms become singular when the cosmological constant tends to zero,
and, does not allow of the flat space limit \footnote{In the case of
massive higher spin fields the mass plays the role of the
dimensionful constant which, similar to $\Lambda$, can be used to
construct electromagnetic and gravitational interactions of the
massive higher spin fields even in flat space, however, such models
usually suffer causality and unitarity problems}.

Consider now the gauge variation of the interaction term
(\ref{nonl})
\begin{eqnarray}\label{varia}
\delta (\triangle G_{m_1m_2})&={{2i}\over
\Lambda}(R_{pm_1m_2q}+R_{pm_2m_1q})\,\gamma_n\nabla^n\nabla^{p}\xi^{q}\nonumber\\
&={i\over
\Lambda}(R_{pm_1m_2q}+R_{pm_2m_1q})\,\gamma_n[\nabla^n,\nabla^{p}]\xi^q+\cdots\,
\end{eqnarray}
where $\cdots$ stand for the terms with the anticommutator of
$\nabla_n$ which are assumed to be harmless, i.e. can be canceled
by an appropriate modification of the gauge transformations of
fields and/or by adding more cubic interaction terms (with higher
derivatives) in to the action and into the equation of motion. If
we restrict ourselves to the consideration of only the first order
in small gravitational interactions, then in (\ref{varia}) the
commutator of derivatives should be restricted to the zero order
contribution of the AdS curvature (\ref{RAdS}), (\ref{comv}). The
AdS curvature (\ref{RAdS}) is proportional to the cosmological
constant which cancels that in the denominator of (\ref{varia}).
So in this approximation the form of the variation of the
non--minimal interaction term (\ref{vari}) reduces to that of the
rest of the field equation (\ref{variR}) with the opposite sign
and thus cancels the latter.

\subsection{Towards a complete non--linear higher spin field
theory}

It turns out that beyond the linear approximation of gravitational
fluctuations the situation with gauge invariance becomes much more
complicated. As the analysis carried out by different people
showed \cite{bengtsson&al,Berends&al,fradkin&vasiliev,deserD}, a
gauge invariant interacting theory of massless higher spin fields
should contain
\begin{itemize}
\item
infinite number of fields of increasing spins involved in the
interaction and in symmetry transformations and
\item
terms with higher derivatives of fields both in the action and in
gauge transformations.
\end{itemize}

No complete action has been constructed so far to describe such an
interacting  theory of infinite number of higher spin fields,
though generic non--linear equations of motion describing higher
spin interactions do exist \cite{vas,progress}. A main problem is
in finding and understanding the (non--abelian) algebraic
structure of the gauge transformations of the higher spin fields
modified by their interactions. In other words, the question is
what is the gauge symmetry algebra which governs the interacting
higher spin theory? Note that in the case of Yang--Mills vector
fields and gravity the knowledge of the structure of non--abelian
gauge symmetries and general coordinate invariance was crucial for
the construction of the complete non--linear actions for these
fields.

To deal simultaneously with the whole infinite tower of higher
spins and to analyze their symmetry and geometrical properties,
one may try to cast them into a finite number of `hyperfields' by
extending space--time with additional directions associated with
infinitely many spin degrees of freedom. In the formulation
discussed above this can be done by introducing auxiliary vector
coordinates $y^m$ \cite{fronsdal78}.

Consider in a $(D+D)$-dimensional space parametrized by $x^m$ and
$y^n$ a scalar field $\Phi(x,y)$ which is analytic in $y^n$. Then
$\Phi(x,y)$ can be presented as a series expansion in powers of
$y^n$
\begin{equation}\label{xyb0}
\Phi(x,y)=\phi(x)+\phi_m(x)\,y^m+\phi_{m_1m_2}(x)\,y^{m_1}y^{m_2}
+\sum_{s=3}^\infty\,\phi_{m_1\cdots m_s}(x)\,y^{m_1}\cdots
y^{m_s}\,.
\end{equation}
We see that the components of this expansion are D-dimensional
symmetric tensor fields $\phi_{m_1\cdots m_s}(x)$. We would like
$\Phi(x,y)$ to have symmetry properties and to satisfy field
equations which would produce the gauge transformations
(\ref{hsg}), the traceless conditions (\ref{tl}), (\ref{dtr}) and
the equations of motion (\ref{be}) of the higher spin fields
$\phi_{m_1\cdots m_s}(x)$. It is not hard to check that the gauge
transformation of the hyperfield $\Phi(x,y)$ should have the form
\begin{equation}\label{xyg}
\delta \Phi(x,y)=y^m\,\partial_m\,\Xi(x,y),
\end{equation}
where as above $\partial_m={\partial\over{\partial x^m}}$, and
higher components of the gauge parameter
$\Xi(x,y)=\sum_{s=0}^\infty\,\xi_{m_1\cdots m_s}(x)y^{m_1}\cdots
y^{m_s}$ are traceless, which is ensured by imposing the condition
\begin{equation}\label{xytl}
\partial^2_y\,\Xi(x,y)=0, \qquad   \partial^2_y\equiv
\eta^{mn}{\partial\over{\partial y^m}}{\partial\over{\partial
y^n}}
\end{equation}
The double tracelessness (\ref{dtr}) is encoded in the equation
\begin{equation}\label{xydtr}
\partial^2_y\,\partial^2_y \,\Phi(x,y)=0,
\end{equation}
and the higher spin field equations (\ref{be}) are derived from
the following equations of motion of the hyperfield $\Phi(x,y)$
\begin{equation}\label{xybe}
\left[\eta^{mn}-y^m{\partial\over{\partial y_n}}+y^m
y^n\partial^2_y \right]\partial_m\partial_n\,\Phi(x,y)=0\,.
\end{equation}

Analogously, to describe the fields with half integer spins let us
introduce a spinorial hyperfield
\begin{equation}\label{xyf0}
\Psi^\alpha(x,y)=\psi^\alpha(x)+\psi^\alpha_m(x)\,y^m+
\sum_{s-{5\over 2} }^\infty\,\psi^\alpha_{{m_1}\cdots m_{s-{1\over
2}}}(x)\,y^{m_1}\cdots y^{m_{s-{1\over 2}}}\,.
\end{equation}
The gauge transformations of $\Psi^\alpha(x,y)$ are
\begin{equation}\label{xygf}
\delta \Psi^\alpha(x,y)=y^m\,\partial_m\,\Xi^\alpha(x,y), \quad
\gamma^m{\partial\over{\partial y^m}} \Xi^\alpha(x,y)=0.
\end{equation}
$\Psi^\alpha(x,y)$ satisfies the `triple' gamma--traceless
condition
\begin{equation}\label{xydtrf}
\gamma^m{\partial\over{\partial y^m}} \partial^2_y\, \Psi(x,y)=0,
\end{equation}
and the equations of motion
\begin{equation}\label{xyfe}
\gamma^m\left[\partial_m-y^n{\partial\over{\partial y^m}}
\partial_n\right]\,\Psi(x,y)=0\,.
\end{equation}
The equations (\ref{xygf}), (\ref{xydtrf}) and (\ref{xyfe})
comprise those for the half--integer spin fields.

The above construction is a simple example of how one can
formulate the free theory of infinite number of higher spin fields
in terms of a finite number of fields propagating in extended
space. In contrast to the Fronsdal formulation, this construction
is on the mass shell. It is not clear how to construct an action
in the extended space which would produce the equations
(\ref{xybe}) and (\ref{xyfe}), neither how to generalize these
equations to include non--linear terms.

Much more sophisticated on--shell formulations which involve
either vector or spinor auxiliary variables and are based on a
solid group--theoretical ground have been developed in
\cite{vas,arbitraryD}.

For instance, to find and study the algebraic and geometrical
structure of higher spin symmetries (at least in $D=4$ and $D=6$),
an alternative description of the higher spin fields  has proved
to be useful. It has been mainly developed by Vasiliev with
collaborators (see \cite{vas}, \cite{progress}  and references
therein) and by Sezgin and Sundell \cite{ss}. This is a so called
unfolded formulation of the equations of motion of higher spin
field theory with the use of spin--tensor representations of the
Lorentz group and auxiliary commuting spinor coordinates.
``Unfolded'' basically means that all fields (including scalars
and spinors) enter into the game with their descendants, i.e.
auxiliary fields which on the mass shell are higher derivatives of
the physical fields. From the algebraic point of view the unfolded
formulation is a particular realization and an extension of a so
called free differential algebra which is also a basis of the
group manifold approach \cite{freeda}. The field equations are
formulated as a zero curvature condition which requires also
0--forms to be involved into the description of systems with
infinite number of degrees of freedom.

\subsection{Unfolded field dynamics} In the unfolded formulation
\cite{progress} the fields of spin $s\geq 1$ in $D=4$ are
described by a generalized vielbein and connection one--form
\begin{equation}\label{omega}
\omega(x,Y)=\sum_{n,p=0}^\infty dx^m\omega_m^{A_1\cdots A_n,\dot
A_1\cdots \dot A_p}(x)\,y_{A_1}\cdots y_{A_p}\,{\bar y}_{\dot
A_1}\cdots {\bar y}_{\dot A_p}
\end{equation}
and by its curvature two--form
\begin{equation}\label{curve}
R(x,Y)=d\omega(x,Y)- (\omega \wedge \star \,\omega)(x,Y),
\end{equation}
where $ (y_A,{\bar y}_{\dot A})=Y_\alpha$ $(A,\dot A=1,2)$ are
auxiliary two--component Weyl spinor variables with even Grassmann
parity which resemble twistors and satisfy the oscillator (or
Moyal star--product) commutation relations
\begin{equation}\label{*}
y_A\star y_B-y_B\star y_A=\epsilon_{AB}, \quad \bar y_{\dot
A}\star \bar y_{\dot B}-\bar y_{\dot B}\star \bar y_{\dot
A}=\epsilon_{\dot A\dot B},
\end{equation}
and the star--product of the connection has been used in the
definition of the curvature (\ref{curve}).

Another object of the unfolded formulation is the zero--form
\begin{equation}\label{C}
C(x,Y)=\sum_{n,p=0} C^{A_1\cdots A_n,\dot A_1\cdots \dot
A_p}(x)\,y_{A_1}\cdots y_{A_n}\,{\bar y}_{\dot A_1}\cdots {\bar
y}_{\dot A_p}
\end{equation}
which contains the scalar field $\phi(x)=C(x,Y)|_{Y=0}$, the
spinor field $\psi^\alpha(x)={\partial\over
Y_\alpha}C(x,Y)|_{Y=0}$ and (Weyl) curvature tensors of the higher
spin fields. The field $C(x,Y)$ is introduced to incorporate the
spin--0 and spin--${1\over 2}$ matter fields, and its higher
components in the $Y$--series expansion are either gauge field
curvature tensors related to (\ref{curve}) or higher derivatives
of the matter fields and of the gauge field curvatures.

The higher spin gauge transformations are
\begin{equation}\label{HSo}
\delta\,\omega(x,Y)=d\xi(x,Y)-(\omega\star\xi)(x,Y)+(\xi\star\omega)(x,Y)\,
\end{equation}
\begin{equation}\label{HSC}
\delta\,C(x,Y)=(\xi\star C)(x,Y)-(C\star\tilde\xi)(x,Y),
\end{equation}
where $\tilde\xi=\xi(x,y,-\bar y)$.

In the free (linearized) higher spin theory, the following
relation holds
\begin{eqnarray}\label{RC}
R_{linear}(x,Y)=&\left\{\partial^A\bar\partial^{\dot
B}\omega(x,Y)\wedge\partial_A\bar\partial^{\dot
C}\omega(x,Y)\right\}|_{Y=0}\,\,\,\bar\partial_{\dot
B}\bar\partial_{\dot C}C(x,0,\bar y)\nonumber\\
&+\left\{\partial^A\bar\partial^{\dot
B}\omega(x,Y)\wedge\partial^C\bar\partial_{\dot
B}\omega(x,Y)\right\}|_{Y=0}\,\,\,\partial_{A}\partial_{
C}C(x,y,0)\,,
\end{eqnarray}
and $C(x,Y)$ satisfies the unfolded field equations
\begin{equation}\label{unc}
i\sigma^m_{A\dot A}{\partial\over{\partial
x^m}}C(x,Y)={\partial\over{\partial y^A}}{\partial\over{\partial
{\bar y^{\dot A}}}}C(x,Y)\,,
\end{equation}
which when written in components are equivalent to the free higher
spin field equations considered in the symmetric tensor
formulation. For details on the non--linear generalization of the
unfolded equations (\ref{RC}) and (\ref{unc}) we refer the reader
to \cite{progress} and references therein.

 An advantage of the unfolded formalism is that it allows
one to treat the whole infinite tower of the higher spin fields
simultaneously and provides a compact form of the higher spin
symmetry transformations which form an infinite dimensional
associative Lie (super)algebra. All this is required, as we have
discussed above, for the construction of the consistent
interacting higher spin theory. The action describing the unfolded
dynamics is still to be found though.

\section{Other developments}
\subsection{Higher spin field theory from dynamics in tensorial
spaces. An alternative to Kaluza--Klein.} The experience of
studying various field theories teaches us that in many cases a
new insight into their structure can be gained by finding and
analyzing a classical dynamical object whose quantization would
reproduce the field theory of interest. The well known examples
are various spinning particle and superparticle models whose
quantum dynamics is described by a corresponding (supersymmetric)
field theory. In all the conventional cases only finite number of
states of different spins can be produced by quantizing particle
models. But, as we have mentioned, for a consistent interacting
higher spin field theory we need an infinite number of states.
These are produced by strings, but as it has been noted, the
associated string field theory is rather complicated. It seems
desirable at first to find and analyze a simpler model with an
infinite number of quantum higher spin states.

Such a superparticle model does exist \cite{bl}. In addition to
the relation to higher spins, this model reveals other interesting
features, such as the invariance under supersymmetry with
tensorial charges (which are usually associated with brane
solutions of Superstring and M--Theory), and it has been the first
example of a dynamical BPS system which preserves more than one
half supersymmetry of the bulk. The study of these features was a
main motivation for the original paper \cite{bl}. BPS states
preserving ${{2n-1}\over{2n}}$ supersymmetries (with $n=16$ for
$D=10,11$) have later on been shown to be building blocks of any
BPS state and conjectured to be hypothetical constituents or
`preons' of M-theory \cite{preons}. The relation of the model of
\cite{bl} to the theory of massless higher spin fields in the
unfolded formulation (\ref{unc}) was assumed in \cite{exotic},
where the quantum states of the superparticle was shown to form an
infinite tower of the massless higher spin fields. This relation
has been analyzed in detail in \cite{misha,misha1,mps}.

Probably, the first person who suggested a physical application of
tensorial spaces to the theory of higher spins was C.~Fronsdal.

 In his Essay of 1985 \cite{fronsdal1} Fronsdal conjectured that
four--dimensional higher spin field theory can be realized as a
field theory on a ten--dimensional tensorial manifold parametrized
by the coordinates
\begin{equation}\label{x}
x^{\alpha\beta}=x^{\beta\alpha}={1\over
2}x^m\gamma_m^{\alpha\beta}+{1\over 4}
y^{mn}\gamma_{mn}^{\alpha\beta}, \quad m,n=0,1,2,3\,; \quad
\alpha,\beta=1,2,3,4\,,
\end{equation}
where $x^m$ are associated with four coordinates of the conventional
$D=4$ space--time and six tensorial coordinates
$y^{mn}=-y^{mn}$ describe spinning
degrees of freedom.

The assumption was that by analogy with, for example, $D=10$ or
$D=11$ supergravities, which are relatively simple theories but
whose dimensional reduction to four dimensions produces very
complicated extended supergravities, there may exist a theory in
ten--dimensional tensorial space whose alternative Kaluza--Klein
reduction may lead in $D=4$ to an infinite tower of fields with
increasing spins instead of the infinite tower of Kaluza--Klein
particles of increasing mass. The assertion was based on the
argument that the symmetry group of the theory should be
$OSp(1|8)\supset SU(2,2)$, which contains the $D=4$ conformal
group as a subgroup such that an irreducible (oscillator)
representation of $OSp(1|8)$ contains each and every massless
higher spin representation of $SU(2,2)$ only once. So the idea was
that using a single representation of $OSp(1|8)$ in the
ten-dimensional tensorial space one could describe  an infinite
tower of higher spin fields in $D=4$ space--time in a simpler way.
Fronsdal regarded the tensorial space as a space on which $Sp(8)$
acts like a group of generalized conformal transformations. Ten is
the minimal dimension of such a space which can contain D=4
space--time as a subspace. For some reason Fronsdal gave only a
general definition and did not identify this ten--dimensional
space with any conventional manifolds, like the ones mentioned
above.

In his Essay Fronsdal also stressed the importance of $OSp(1|2n)$
supergroups for the description of theories with superconformal
symmetry. In the same period and later on different people studied
$OSp(1|2n)$ supergroups in various physical contexts. For
instance, $OSp(1|32)$ and $OSp(1|64)$ have been assumed to be
underlying superconformal symmetries of string- and M-theory.

The tensorial superparticle model of Bandos and Lukierski
\cite{bl} turned out to be the first dynamical realization of the
Fronsdal proposal.

The tensorial particle action has the following form
\begin{equation}\label{action}
S[X,\lambda]=\int\,
E^{\alpha\beta}\left(X(\tau)\right)\,\lambda_\alpha(\tau)\,\lambda_\beta(\tau),
\end{equation}
where $\lambda_\alpha(\tau)$ is an auxiliary commuting real
spinor, a \textit{twistor--like} variable, and
$E^{\alpha\beta}(x(\tau))$ is the pull back on the particle
worldline of the tensorial space vielbein. In flat tensorial space
\begin{equation}\label{Omega}
E^{\alpha\beta}(X(\tau))=d\tau\,\partial_\tau
X^{\alpha\beta}\,(\tau)=dX^{\alpha\beta}\,(\tau)\,.
\end{equation}
The dynamics of particles on the supergroup manifolds $OSp(N|n)$
(which are the tensorial extensions of AdS superspaces) was
considered for $N=1$ in \cite{preit,mps} and for a generic $N$ in
\cite{misha,misha1}. The twistor--like superparticle in $n=32$
tensorial superspace was considered in \cite{Bandos:2003us} as a
point--like model for BPS preons \cite{preons}, the hypothetical
${{31}\over{32}}$--supersymmetric constituents of M--theory.

The action (\ref{action}) is manifestly invariant under global
$GL(n)$ transformations. Without going into details which the
reader may find in \cite{bl,misha,mps}, let us note that the
action (\ref{action}) is invariant under global $Sp(2n)$
transformations, acting non--linearly on $X^{\alpha\beta}$ and on
$\lambda_\alpha$, {\it i.e.} it possesses the symmetry considered
by Fronsdal to be an underlying symmetry of higher spin field
theory in the case $n=4$, $D=4$ \cite{fronsdal1}.

Applying the Hamiltonian analysis to the particle model described
by (\ref{action}) and (\ref{Omega}), one finds that the momentum
conjugate to $X^{\alpha\beta}$ is related to the twistor--like
variable $\lambda_\alpha$ via the constraint
\begin{equation}\label{Penroselike}
P_{\alpha\beta}=\lambda_\alpha\lambda_\beta\,.
\end{equation}
This expression is the direct analog and generalization of the
Cartan--Penrose (twistor) relation for the particle momentum
$P_m=\lambda\gamma_m\lambda$. In virtue of the Fierz identity
\hbox {$\gamma_{m(\alpha\beta}\,\gamma^m_{\gamma)\delta}=0$} held
in $D=3,4,6$ and $10$ space--time, the twistor particle momentum
is light--like in these dimensions. Therefore, in the tensorial
spaces corresponding to these dimensions of space--time the
first--quantized particles are massless \cite{bl,exotic}.

The quantum counterpart of (\ref{Penroselike}) is the equation
\cite{exotic}
\begin{equation}\label{l}
D_{\alpha\beta}\Phi(X,\lambda)=\left({\partial\over{\partial
X^{\alpha\beta}}}-i\lambda_\alpha\lambda_\beta\right)\Phi(X,\lambda)=0\,,
\end{equation}
where the wave function $\Phi(X,\lambda)$ depends on
$X^{\alpha\beta}$ and $\lambda_\alpha$.  The general solution of
(\ref{l}) is the plane wave
\begin{equation}\label{soll}
\Phi(X,\lambda)=e^{iX^{\alpha\beta}\lambda_\alpha\lambda_\beta}\varphi(\lambda),
\end{equation}
where $\varphi(\lambda)$ is a generic function of
$\lambda_\alpha$.

One can now Fourier transform the function (\ref{soll}) to another
representation
\begin{equation}\label{yr}
C(X,Y)=\int\,d^4\lambda\,e^{-iY^\alpha\lambda_\alpha}\Phi(X,\lambda)=\int\,d^4\lambda\,e^{-iY^\alpha\lambda_\alpha+
i X^{\alpha\beta}\lambda_\alpha\lambda_\beta}\varphi(\lambda).
\end{equation}
The wave function $C(X,Y)$ satisfies the Fourier transformed
equation
\begin{equation}\label{Y}
\left({\partial\over{\partial
X^{\alpha\beta}}}+i{\partial^2\over{\partial Y^\alpha\partial
Y^\beta}}\right)C(x,Y)=0\,,
\end{equation}
which is similar to the unfolded equation (\ref{unc}) and which
actually reduces to the latter \cite{misha,mps}.

Quantum states of the tensorial superparticle satisfying eq.
(\ref{l}) was shown
 to form an infinite series of massless higher spin states in
 $D=4,6$ and 10  space--time
\cite{exotic}. In \cite{preit} quantum superparticle dynamics on
$OSp(1|4)$ was assumed to describe higher spin field theory in
$N=1$ super $AdS_4$.

In \cite{exotic} it was shown explicitly how the alternative
Kaluza--Klein compactification produces higher spin fields. It
turns out that in the tensorial superparticle model, in contrast
to the conventional Kaluza--Klein theory, the compactification
occurs in the momentum space and not in the coordinate space.  The
coordinates conjugate to the compactified momenta take discrete
(integer and half integer values) and describe spin degrees of
freedom of the quantized states of the superparticle in
conventional space--time.

In \cite{misha} M. Vasiliev has extensively developed this subject
by having shown that the first--quantized field equations (\ref{Y})
in tensorial superspace of a bosonic dimension
${n(n+1)}\over 2$ and of a fermionic dimension $nN$ are
$OSp(N|2n)$ invariant, and for $n=4$ correspond to the unfolded
higher spin field equations in $D=4$. It has also been shown
\cite{misha1} that the theory possesses properties of causality
and locality.

As was realized in \cite{misha,misha1}, the field theory of
quantum states of the tensorial particle is basically a classical
theory of two fields in the tensorial space, a scalar field
$b(X^{\alpha\beta})$ and a spinor field
$f^\alpha(X^{\beta\gamma})$. These fields form a fundamental
linear representation of the group $OSp(1|2n)$ and satisfy the
following tensorial equations
\begin{equation}\label{bf}
(\partial_{\alpha\beta}\partial_{\gamma\delta}-\partial_{\alpha\gamma}\partial_{\beta\delta})b(X)=0,
\quad \partial_{\alpha\beta} f_\gamma(X)-\partial_{\alpha\gamma}
f_\beta(X)=0\,.
\end{equation}
In the case of $n=4$ (\ref{x}) the fields $b(X)$ and $f_\alpha(X)$
subject to eqs. (\ref{bf})  describe the infinite tower of the
massless (conformally invariant) fields of all possible integer
and half--integer spins in the physical four--dimensional subspace
of the ten--dimensional tensorial space \cite{fronsdal1,misha}. In
the cases of $n=8$ and $n=16$ which correspond to $D=6$ and $D=10$
space--time, respectively, the equations (\ref{bf}) describe
conformally invariant higher spin fields with self--dual field
strengths \cite{D4610}.

 Let us consider in more detail the case of $n=4$ and $D=4$ we split $X^{\alpha\beta}$ onto
$x^m$ and $y^{mn}$ as in eq. (\ref{x}), the system of equations
(\ref{bf}) takes the form
\begin{eqnarray}\label{xyb}
&\partial_p\,\partial^p\,b(x^l,y^{mn})=0, \quad
\partial_p\,\partial_q\,b(x^l,y^{mn})
-4\partial_{pr}\,\partial^r_{~q}\,b(x^l,y^{mn})=0,\quad
\partial^{~p}_{q}\,\partial_p\,\,b(x^l,y^{mn})=0, \quad \nonumber\\
& \epsilon^{pqrt}\partial_q\,\partial_{rt}\,\,b(x^l,y^{mn})=0,
\quad
\epsilon^{pqrt}\partial_{pq}\,\partial_{rt}\,\,b(x^l,y^{mn})=0,
\end{eqnarray}
\begin{equation}\label{xyf}
\gamma^p\,\partial_{p}\,f(x^l,y^{mn})=0, \quad
\left[{\partial_p}-2\gamma^r\,{\partial_{rp}}\right]f(x^l,y^{mn})=0\,,
\end{equation}
where $\partial_p$ and $\partial_{rp}$ are the derivatives along
$x^p$ and $y^{rp}$, respectively.

Then let us expand $b(x,\,y)$ and $f_\alpha(x,y)$ in series of
$y^{mn}$
\begin{eqnarray}\label{ymn}
b(x^l,\,y^{mn})&=\phi(x)+y^{m_1n_1}F_{m_1n_1}(x)
+y^{m_1n_1}\,y^{m_2n_2}\,[R_{m_1n_1,m_2n_2}(x)-{1\over 2}\eta_{m_1m_2}\partial_{n_1n_2}\phi(x)]\nonumber\\
&+\sum_{s=3}^{\infty}\,y^{m_1n_1}\cdots
y^{m_sn_s}\,[R_{m_1n_1,\cdots,m_sn_s}(x)+\cdots]\,,\nonumber\\
~&~\\
 f^\alpha(x^l,y^{mn})
 &{\hspace{-150pt}}
=\psi^\alpha(x)+y^{m_1n_1}[{\cal R}^\alpha_{m_1n_1}(x)-{1\over
2}\partial_{m_1}(\gamma_{n_1}\psi)^\alpha]\nonumber\\
& +\sum_{s={5\over 2}}^{\infty}\,y^{m_1n_1}\cdots y^{m_{s-{1\over
2}}n_{s-{1\over 2}}}\,[{\cal R}^\alpha_{m_1n_1,\cdots,m_{s-{1\over
2}}n_{s-{1\over 2}}}(x)+\cdots]\,.\nonumber
\end{eqnarray}
In (\ref{ymn}) $\phi(x)$ and $\psi^\alpha(x)$ are scalar and spin
1/2 field, $F_{m_1n_1}(x)$ is the Maxwell field strength,
$R_{m_1n_1,m_2n_2}(x)$ is the curvature tensor of the linearized
gravity, ${\cal R}^\alpha_{m_1n_1}(x)$ is the Rarita--Schwinger
field strength and other terms in the series stand for generalized
Riemann curvatures of spin-$s$ fields (which also contain
contributions of derivatives of lower spin fields denoted by dots,
as in the case of the Rarita--Schwinger field and gravity). The
scalar and the spinor field satisfy, respectively, the
Klein--Gordon and the Dirac equation,  and the higher spin field
curvatures satisfy the Bianchi identities (\ref{B1}), (\ref{B2})
and the linearized higher spin field equations (\ref{trbis}) and
(\ref{trfis}) in $D=4$ space--time. Similar equations also follow
from the unfolded equations (\ref{unc}). In the model under
consideration they are consequences of the field equations
(\ref{bf}), or equivalently of (\ref{xyb}) and (\ref{xyf}) in the
flat tensorial space. The generalization of the equations
(\ref{bf}) to a field theory on the tensorial manifold $OSp(1|n)$,
which for $n=4$ corresponds to the theory of higher spin fields in
$AdS_4$, has been derived in \cite{mps}.

An interesting and important problem is to find a simple and
appropriate non--linear generalization of equations (ref{bf} which
would correspond to an interacting higher spin field theory. An
attempt to construct such a generalization in the framework of
tensorial superspace supergravity was undertaken in \cite{bpst}.

\subsection{Massless higher spin field theory as a tensionless
limit of superstring theory}

In these lectures we have considered the formulations of massless
higher spin field theory which are not directly related to String
Theory. A natural question arises which formulation one can derive
from String Theory at the tensionless limit $T\sim {1\over
\alpha'} \rightarrow 0$. This has been a subject of a number of
papers \cite{ht}--\cite{am} (and references therein) which we
briefly sketch below.

Consider, for instance a free open bosonic string in flat
space--time, whose worldsheet is parametrized by a `spatial'
coordinate $\sigma\in[0,\pi]$  and a `time' coordinate $\tau$.
String dynamics is described by the coordinates
\begin{equation}\label{sx}
X^m(\tau,\sigma)=x^m+2\alpha' p^m\tau +i\sqrt{2\alpha'}\,\sum_{n
\not = 0}^\infty {1\over n}a^m_n e^{-in\tau}cos(n\sigma)
\end{equation}
and momenta
\begin{equation}\label{px}
P^m(\tau,\sigma)=p^m +{1\over{\sqrt{2\alpha'}}}\,\sum_{n\not =
0}^\infty a^m_n e^{-in\tau}cos(n\sigma),
\end{equation}
where $x^m$ and $p^m$ are the center of mass variables and $a^m_n$
are the string oscillator modes satisfying (upon quantization) the
commutation relations $[p^m,x^p]=-i\eta^{mp}$,
$[a^m_n,a^p_l]=n\delta_{n+l}\eta^{mp}$.

String dynamics is subject to the Virasoro constraints
\begin{equation}\label{Lk}
L_k={1\over 2}\sum_{n=-\infty}^{+\infty}
a^m_{k-n}a_{m\,n}=\sqrt{2\alpha'}\,p^m\,a_{m\,k}+{1\over
2}\sum_{n\not=k,0} a^m_{k-n}a_{m\,n}, \quad k\not = 0,
\end{equation}
\begin{equation}\label{L0}
L_0=2\alpha' p^mp_m+\sum_{n>0}a^m_{-n}a_{m\,n}.
\end{equation}
The latter produces the mass shell condition for the string states
\begin{equation}\label{ms}
M^2=-p^mp_m={1\over {2\alpha'}}\sum_{n>0}a^m_{-n}a_{m\,n}\,.
\end{equation}
We observe that in the tensionless limit $\alpha' \rightarrow
\infty$ all string states become massless, while the properly
rescaled Virasoro constraints become at most linear in the
oscillator modes
\begin{equation}\label{0}
l_0={1\over {2\alpha'}}L_0|_{\alpha' \rightarrow \infty}=p^mp_m,
\qquad l_k={1\over{\sqrt{2\alpha'}}}L_k|_{\alpha' \rightarrow
\infty}=p_m\,a^m_{k}
\end{equation}
and satisfy a simple algebra without any central charge
\begin{equation}\label{lalgebra}
[l_0,l_k]=0, \qquad [l_j,l_k]=\delta_{j+k}\,l_0\,.
\end{equation}
Thus in the tensionless limit the quantum consistency of string
theory does not require any critical dimension for the string to
live in. Note that at $\alpha'\rightarrow \infty$ the string
coordinate (\ref{sx}) blows up and is not well defined, while the
oscillator modes remain appropriate variables for carrying out the
quantization of the theory.

The corresponding nilpotent BRST charge takes the form
\begin{equation}\label{BRST}
Q=\sum_{n=-\infty}^{+\infty}(c_{-n}\,l_n-{n\over
2}b_0\,c_{-n}c_n)\,,
\end{equation}
where $c_n$ and $b_n$ are the ghosts and anti--ghosts associated
with the constraint algebra (\ref{lalgebra}).

The BRST charge can be used to construct a free action for the
string field states $|\Phi>$, obtained by acting on the Fock
vacuum by the creating operators,
\begin{equation}\label{fsa}
S={1\over 2}\int\, <\Phi| Q|\Phi>\,.
\end{equation}
The action  (\ref{fsa}) can be used for the derivation of a
corresponding action and equations of motions of the higher spin
fields encoded in $|\Phi>$. As has been shown in \cite{am} such an
action and equations of motion are more involved than eqs.
(\ref{ba}), (\ref{fa}), (\ref{be}) and (\ref{fe}) since they
contain intertwined (triplet) fields of different spins. The
equations (\ref{ba}), (\ref{fa}), (\ref{be}) and (\ref{fe}) are
obtained from (\ref{fsa}) upon gauge fixing part of the available
local symmetry and by eliminating auxiliary fields.

Further details the interested reader can find in \cite{giulio,am}
and references therein. We should note that the tensionless limit
of the string considered here differs from the so called null
string models \cite{null}. In these models, in contrast to the way
of getting the tensionless string discussed above the limit is
taken in such a way that the string coordinate $X^m(\sigma,\tau)$
remains a well defined variable, while the oscillator modes
disappear. As a result the quantum states of the null strings
correspond to a continuous set of massless particles without
(higher) spin.

\section{Conclusion}
In these lectures we have described main features and problems of
higher spin field theory and have flashed some ways along which it
has been developed over last years.

\section*{Acknowledgments}
  I am grateful to Igor Bandos, Jose de
Azc\'arraga, Alexander Filippov, Jerzy Lukierski, Paolo Pasti,
Mario Tonin and attenders of the courses and seminars for the
encouragement to write these lecture notes. I would also like to
thank Misha Vasiliev for useful comments. This work was partially
supported by the Grant N 383 of the Ukrainian State Fund for
Fundamental Research, by the INTAS Research Project N 2000-254, by
the European Community's Human Potential Programme under contract
HPRN-CT-2000-00131 Quantum Spacetime, by the EU
MRTN-CT-2004-005104 grant `Forces Universe', and by the MIUR
contract no. 2003023852.

\end{document}